\documentclass[fleqn,usenatbib]{mnras}

\usepackage{newtxtext,newtxmath}
\usepackage[T1]{fontenc}
\usepackage{ae,aecompl}
\usepackage{lipsum}

\usepackage{graphicx}	
\usepackage{amsmath}	
\usepackage{amssymb}	
\def\Dp{DUSTGRAIN-\textit{pathfinder}}
\def\wf{w_{\mathrm{f}}}
\def\wc{w_{\mathrm{m}}}
\def\wnd{\texttt{wnd-charm}}
\newcommand{\dust}{{\small DUSTGRAIN} }
\newcommand{\dustp}{{\small DUSTGRAIN}-\emph{pathfinder} }

\newcommand{\mapsim}{\textsc{MapSim} }
\newcommand{\lcdm}{$\mathrm{\Lambda CDM}$ }
\newcommand{\zstwo}{$z_{\mathrm{s}}=2$ }
\newcommand{\zsone}{$z_{\mathrm{s}}=1$ }
\newcommand{\zspfive}{$z_{\mathrm{s}}=0.5$ }

\title[ML @ $\nu$MG]{On the dissection of degenerate cosmologies with machine learning}

\author[J. Merten et al.]{Julian Merten,$^{1}$\thanks{E-mail: julian.merten@inaf.it}
Carlo Giocoli,$^{1,2,3,4}$
Marco Baldi,$^{1,3,4}$
Massimo Meneghetti,$^{1,3,4}$ \newauthor
Austin Peel,$^{5}$
Florian Lalande,$^{5,6}$
Jean-Luc Starck,$^{5}$
and Valeria Pettorino$^{5}$
\\
$^{1}$INAF--Osservatorio di Astrofisica e Scienza dello Spazio di Bologna, Via Gobetti 93/3, 40129, Bologna, Italy\\
$^{2}$Dipartimento di Fisica e Scienze della Terra, Universit\`a degli Studi di Ferrara, via Saragat 1, I-44122 Ferrara, Italy\\
$^{3}$Dipartimento di Fisica e Astronomia, Alma Mater Studiorum Universit\`a di Bologna, via Gobetti 93/2, I-40129 Bologna, Italy\\
$^{4}$INFN -- Sezione di Bologna, viale Berti Pichat 6/2, I-40127 Bologna, Italy\\
$^{5}$AIM, CEA, CNRS, Universit\'{e} Paris-Saclay, Universit\'{e} Paris Diderot, Sorbonne Paris Cit\'{e}, F-91191 Gif-sur-Yvette, France\\
$^{6}$ENSAI, rue Blaise Pascal, 35170 Bruz, France
}

\date{\textcopyright~2018 INAF. All rights reserved.\\
  Accepted for publication by the Monthly Notices of the Astronomical 
Society.}

\pubyear{2019}

\begin{document}
\label{firstpage}
\pagerange{\pageref{firstpage}--\pageref{lastpage}}
\maketitle

\begin{abstract}
Based on the \Dp~suite of simulations, we investigate observational degeneracies between nine models of modified gravity and massive neutrinos. Three types of machine learning techniques are tested for their ability to discriminate lensing convergence maps by extracting dimensional reduced representations of the data. Classical map descriptors such as the power spectrum, peak counts and Minkowski functionals are combined into a joint feature vector and compared to the descriptors and statistics that are common to the field of digital image processing. To learn new features directly from the data we use a Convolutional Neural Network (CNN). For the mapping between feature vectors and the predictions of their underlying model, we implement two different classifiers; one based on a nearest-neighbour search and one that is based on a fully connected neural network. \\
We find that the neural network provides a much more robust classification than the nearest-neighbour approach and that the CNN provides the most discriminating representation of the data. It achieves the cleanest separation between the different models and the highest classification success rate of 59\% for a single source redshift.
Once we perform a tomographic CNN analysis, the total classification accuracy increases significantly to 76\% with no observational  degeneracies remaining. Visualising the filter responses of the CNN at different network depths provides us with the unique opportunity to learn from very complex models and to understand better why they perform so well. 
\end{abstract}




\section{Introduction}
\label{SEC::INTRO}


The standard $\Lambda $CDM cosmological model -- based on a cosmological constant as the source of the observed accelerated cosmic expansion \citep[][]{Riess_etal_1998,Perlmutter_etal_1999,Schmidt_etal_1998} and on cold dark matter particles as the bulk of the clustering mass in the universe \citep[][]{white78,white93,white96b,springel05b} --  has survived the past two decades of cosmological observations targeted to a wide range of independent probes. This includes the statistical properties of Cosmic Microwave Background (CMB) anisotropies \citep[][]{wmap9,Planck_2018_VI}, the large-scale distribution and dynamics of visible galaxies \citep[][]{Parkinson_etal_2012,SDSS-III-final,Pezzotta_etal_2017}, weak gravitational lensing signals \citep[][]{Fu_etal_2008,Joudaki2017,Hildebrandt_etal_2017,Troxel2018,Hikage2018}, the abundance of galaxy clusters, as well as its time evolution \citep{Vikhlinin2009,Planck2016}. 

Despite this astonishing success, the fundamental nature of the two main
ingredients of the $\Lambda $CDM model -- summing up to about $95\%$ of the 
total energy density of the universe -- remains unknown. On one side, the energy 
scale associated with the cosmological constant does not find any reasonable  
explanation in the context of fundamental physics, with predictions based on the 
standard model of particle physics failing by tens of orders of magnitude. On 
the other hand, no clear detection -- direct or indirect --  of any new 
fundamental particle that may be associated with cold dark matter has been made  
despite a longstanding chase through astrophysical observations 
\citep[][]{Aartsen_etal_2013,Ackermann_etal_2017,Albert_etal_2017} and 
laboratory experiments \citep[see 
e.g.][]{Bernabei2018,ATLAS_WIMP_2013,CMS_WIMP_2016}.

This leaves the next generation of cosmological observations with the arduous challenge of clarifying the fundamental nature of the dark sector by systematically scrutinising the huge wealth of high-quality data that will be made available in the near future by several wide-field surveys \citep[such as][]{euclidredbook,LSST,WFIRST,J-PAS}. As a matter of fact, any possible insights from future datasets must come in the form of very small deviations from the expectations of the $\Lambda $CDM model, otherwise past observations would have already detected them. This suggests that either the fundamental physics behind dark energy and dark matter is indeed extremely close to that of General Relativity (GR) with a cosmological constant and heavy fundamental particles with negligible thermal velocities, respectively, or that a more radical shift from this standard paradigm is hidden and masked by  other effects such as an observational degeneracy with some not yet fully constrained cosmological parameter. The latter scenario may result in a severe limitation of the discriminating power of observations, thereby providing a particularly challenging testbed for the next generation of cosmological surveys. 

A typical example of such a possible intriguing situation is given by the well-known degeneracy between some  Modified Gravity \citep[MG, see e.g.][for a recent review on a wide range of MG scenarios]{Euclid_TWG_2} theories and the yet unknown value of the neutrino mass. It is now generally accepted \citep[][]{Baldi_etal_2014,He_2013,Motohashi_etal_2013,Wright_Winther_Koyama_2017} that MG  theories  such  as $f(R)$  gravity  \citep[][]{Buchdahl_1970} are strongly  observationally degenerate with the  effects of massive neutrinos on structure formation \citep[see][]{Baldi_etal_2014}.  Some commonly adopted statistics  such  as  the  matter  auto-power  spectrum \citep[][]{Giocoli_Baldi_Moscardini_2018}, the  lensing convergence power spectrum \citep[][]{Peel2018},  and the halo mass function \citep[][]{Hagstotz_etal_2018} may hardly distinguish standard  \lcdm  expectations  from  some specific combinations of the $f(R)$  gravity parameters and the total neutrino mass.

As such degeneracies extend down to the non-linear regime of structure formation, the use of full numerical simulations currently represents the only viable method to explore these scenarios, even though alternative approaches based on approximate methods \citep[see e.g.][]{Wright_Winther_Koyama_2017} have been developed in the last years and are being tested and calibrated against simulations. In the present work, we will explore the prospects of using machine learning techniques applied to numerical simulations of both MG and \lcdm cosmologies that are highly observationally degenerate through standard observational statistics.

Several variants of higher-order statistics have been applied in the past to characterise cosmological data sensitive to the late-time evolution of structure in the Universe. Recent analyses of the weak lensing (WL) \citep{Bartelmann2001} data from CFHTLens \citep{Heymans2012} used either higher-order (>2) moments of the convergence field \citep{VanWaerbeke2013a}, or Minkowski functionals \citep{Petri2015} to draw cosmological inference from a data description that goes beyond two-point statistics. \citet{Martinet2018} and \citet{Shan2018} applied peak count statistics \citep{Dietrich2010,Kratochvil2010} to shear and convergence fields from KiDS \citep{Hildebrandt_etal_2017} and \citet{Gruen2018} used counts-in-cells \citep{Friedrich2018} to extract information from the DES \citep{DES2018} catalogues. A new set of techniques based on deep learning \citep{Lecun2015} currently has gained momentum in many scientific fields, including astrophysics. The extremely complex models which can be constructed through a modular building-block concept \citep[e.g.][]{Chollet2017} have been very successful for tasks like language translation \citep[e.g.][]{Wu2016,Johnson2016}, text and handwriting recognition \citep[e.g.][]{Graves2013}, as well as for the classification of images \citep[starting with the seminal work of][]{Krizhevsky2012}. In cosmology, deep learning is used for the extraction of information from N-body simulations \citep{Ravanbakhsh2017}, to learn the connection between initial conditions and the final shape of structure \citep{Lucie-Smith2018}, for the characterisation of point spread functions \citep{Herbel2018} or the measurement of shear for WL \citep{Springer2018}, the characterisation of non-Gaussian structure in mass maps \citep{Gupta2018}, the determination of galaxy cluster X-ray masses \citep{Ntampaka2018} and the fast creation of simulated data using generative adversarial networks \citep{Rodriguez2018}. In this work, we will use such techniques to break the degeneracies between models of modified gravity in the presence of massive neutrinos.

The text is organised as follows: Section \ref{SEC::SIM} gives an overview of the numerical simulations and the creation of the mass maps that constitute our main data set. In section \ref{SEC::METHOD} we introduce the different characterisation and classification techniques that we apply to the mass map data and show the results that they produce in section \ref{SEC::RESULTS}. We present our conclusions in section \ref{SEC::CONCL}. Two appendices provide more details on certain technical aspects of the computer vision (appendix \ref{SEC::APP1}) and deep neural network (appendix \ref{SEC::APP2}) methods we are using.

\section{Numerical simulations}
\label{SEC::SIM}
We perform our analysis on a set of WL maps extracted  from a suite of  cosmological dark matter-only simulations  called the \dustp runs. These simulations represent a preliminary calibration sample for the  \dust (Dark  Universe Simulations to  Test GRAvity  In the presence  of Neutrinos)  project,  an  ongoing  numerical effort aimed  at investigating cosmological models characterised  by a modification of the laws of gravity  from their standard GR form  and by a non-negligible fraction  of the  cosmic matter  density being  made of standard  massive neutrinos.

\subsection{\Dp}
The modification of gravity considered in the \dust project consists in an $f(R)$ model defined by the Action \citep[][]{Buchdahl_1970}
\begin{equation}
\label{fRaction}
  S = \int {\rm d}^4x \, \sqrt{-g} \left( \frac{R+f(R)}{16 \pi G} +
  {\cal L}_m \right).
\end{equation}
We assume a specific analytical form for the $f(R)$ function \citep{Hu_Sawicki_2007}
\begin{equation}
\label{fRHS}
f(R) = -m^2 \frac{c_1 \left(\frac{R}{m^2}\right)^n}{c_2
  \left(\frac{R}{m^2}\right)^n + 1},
\end{equation}
where $R$ is the Ricci scalar curvature, $ m^2  \equiv H_0^2 \Omega_{\rm M}$  is a mass scale,  while $\left\{ c_{1}, c_{2}, n\right\}\geq 0$ are free  parameters of the model. Such a form is particularly popular and widely investigated as it allows one to recover with arbitrary precision a $\Lambda $CDM background expansion history by choosing $c_{1}/c_{2} = 6\Omega _{\Lambda}/\Omega _{\rm  M}$.  Here $\Omega_{\Lambda}$ and  $\Omega_{\rm M}$ are the vacuum and matter energy density, respectively, under the condition  $c_2 (R/m^2)^n  \gg 1$,  so that  the scalar  field $f_{R}$ takes the approximate form
\begin{equation}
  f_R \approx -n \frac{c_1}{c_2^2}\left(\frac{m^2}{R}\right)^{n+1}.
\label{eq:fR-R,n_relation}
\end{equation}
By restricting to the case $n=1$ the only remaining free parameter of the model can be written as
\begin{equation}
f_{R0}\equiv -\frac{1}{c_{2}}\frac{6\Omega _{\Lambda }}{\Omega_{M}}\left( \frac{m^{2}}{R_{0}}\right) ^{2}
\end{equation}
and its absolute value $|f_{R0}|$ will quantify how much the model departs from GR.

The \dustp simulations have been devised to  sample the $\left\{f_{R0},m_{\nu}\right\}$ parameter  space and to identify highly degenerate combinations of parameters. Some analyses of the corresponding WL signal have been presented by \citet{Giocoli_Baldi_Moscardini_2018} and \citet{Peel2018}, while \citet{Hagstotz_etal_2018} have used the simulations to calibrate a theoretical modelling of the halo mass function in $f(R)$ gravity with and without the contribution of massive neutrinos. In this further paper, we will use machine learning techniques to tackle the issue of observational degeneracy in these combined models based on the WL reconstruction described in \citet{Giocoli_Baldi_Moscardini_2018}. A similar approach, focused on a subset of particularly degenerate models is presented in Peel et al. (2018b, PRL submitted).

From a technical point of view, the \dustp runs are cosmological  collisionless  simulations including $768^{3}$ Dark Matter particles  of mass $m_{\rm  CDM}= 8.1\times  10^{10}$  M$_{\odot }/h$  (for  the case  of $m_{\nu  }=0$) and  as  many neutrino  particles (for  the case  of $m_{\nu }>0$) in a $(750 \mathrm{Mpc}/h)^{3}$ cosmological volume with periodic boundary conditions evolving under the  effect of  a gravitational  interaction defined  by equation \ref{fRaction}.  The simulations  have been performed with the  {\small  MG-GADGET}  code \citep[see][]{Puchwein_Baldi_Springel_2013}, a modified version of the {\small  GADGET} code  \citep[][]{gadget-2} that  implements all the modifications that  characterise   $f(R)$ gravity \citep[see][for   more   details on the algorithm]{Puchwein_Baldi_Springel_2013}.    {\small  MG-GADGET} has been extensively tested  \citep[see e.g. the Modified Gravity code  comparison  project  described in  ][]{Winther_etal_2015}  and employed recently for a wide variety of applications 
\citep[][]{Baldi_Villaescusa-Navarro_2018,Arnold_etal_2018,Arnold_Puchwein_Springel_2014,Arnold_Puchwein_Springel_2015,Roncarelli_Baldi_Villaescusa-Navarro_2018,Arnold_Springel_Puchwein_2016,Naik_etal_2018}. For the \dustp simulations, as was already done in \citet{Baldi_etal_2014}, we have combined the {\small MG-GADGET} solver with the particle-based implementation  of massive neutrinos developed by  \citet{Viel_Haehnelt_Springel_2010}. This allowed us to include  massive neutrinos in the simulations as an independent family of particles with its  own initial  transfer  function  and velocity  distribution. Initial  conditions  have been  generated  following  the approach  of e.g.~\citet{Zennaro_etal_2017} and \citet{Villaescusa-Navarro_etal_2018}  at  the  starting  redshift  of  the  simulation  $z_{i}=99$ with thermal neutrino velocities added on top of the gravitational velocities by random  sampling  the  neutrino  momentum distribution  at  the initial redshift.

Standard  cosmological parameters are set  to be consistent      with      the      Planck      $2015$      constraints \citep[][]{Planck_2015_XIII}. Concerning non-standard parameters, the \dustp  simulations spanned the range $-1\times 10^{-4} \le f_{R0} \le -1\times 10^{-6}$  for the scalar amplitude and $0~{\rm eV} \le m_{\nu  } \le 0.3~{\rm eV}$ for the  neutrino mass, for a total of $20$ simulations. In the present work, we will consider a subset of the full \dustp suite consisting of nine simulations whose specifications are summarised in table~\ref{tab:sims}.

\begin{table*}
\begin{tabular}{lccccccc}
Simulation Name & Gravity type  &  
$f_{R0} $ &
$m_{\nu }$ [eV] &
$\Omega _{\rm CDM}$ &
$\Omega _{\nu }$ &
$m^{p}_{\rm CDM}$ [M$_{\odot }/h$] &
$m^{p}_{\nu }$ [M$_{\odot }/h$] \\
\\
\hline
$\Lambda $CDM & GR & -- & 0 & 0.31345 & 0 & $8.1\times 10^{10}$  & 0 \\
$f_{4}$ & $f(R)$  & $-1\times 10^{-4}$ & 0 & 0.31345 & 0 & $8.1\times 10^{10}$  & 0 \\
$f_{5}$ & $f(R)$  & $-1\times 10^{-5}$ & 0 & 0.31345 &0  & $8.1\times 10^{10}$  & 0 \\
$f_{6}$ & $f(R)$  & $-1\times 10^{-6}$ & 0 & 0.31345 & 0 & $8.1\times 10^{10}$  & 0 \\
$f_{4}^{0.3}$ & $f(R)$  & $-1\times 10^{-4}$ & 0.3 & 0.30630 & 0.00715 & $7.92\times 10^{10}$ & $1.85\times 10^{9}$ \\
$f_{5}^{0.15}$ & $f(R)$  & $-1\times 10^{-5}$ & 0.15 & 0.30987 & 0.00358 & $8.01\times 10^{10}$ & $9.25\times 10^{8}$ \\
$f_{5}^{0.1}$ & $f(R)$  & $-1\times 10^{-5}$ & 0.1 & 0.31107 & 0.00238 & $8.04\times 10^{10}$ & $6.16\times 10^{8}$ \\
$f_{6}^{0.1}$ & $f(R)$  & $-1\times 10^{-6}$ & 0.1 & 0.31107 & 0.00238 & $8.04\times 10^{10}$ & $6.16\times 10^{8}$ \\
$f_{6}^{0.06}$ & $f(R)$  & $-1\times 10^{-6}$ & 0.06 & 0.31202 & 0.00143 & $8.07\times 10^{10}$ & $3.7\times 10^{8}$  \\
\hline
\end{tabular}
\caption{The subset of the \dustp  simulations considered in this work with  their   specific  parameters.   $f_{R0}$  represents   the  MG parameter,   $m_\nu$   and   $m^p_{\nu}$  the   neutrino   mass   in electron volts and in $M_{\odot}/h$ as implemented in the simulation, $m^p_{\rm  CDM}$ cold  dark matter  particle mass,  and $\Omega_{\rm CDM}$ and  $\Omega_{\nu}$ the $\mathrm{CDM}$ and  neutrino density parameters, respectively.}
\label{tab:sims}
\end{table*}

\subsection{Lensing light-cones}
\label{SEC::SIM::lightcones}
For all simulations we  stored $34$ snapshots at different redshifts that allow us to construct lensing light-cones  up to a source redshift $z_s=4$ without gaps. Different methods have been developed to produce lensing light-cones from  large cosmological N-body simulations.   Recent  works  have  employed  post-processing  reconstructions based   on  the slicing  of  a set  of comoving particle snapshots \citep[as e.g.  in][]{hilbert08,hilbert09,giocoli16a,shirasaki17}, as well as on-the-fly algorithms  capable of storing  only the projected matter density on a given field-of-view without resorting on the flat-sky approximation  \citep[see e.g.][]{barreira16,Arnold_etal_2018}. In this work we use  the \mapsim routine \citep{giocoli14,tessore15,castro17} which is based on the former strategy.  We use the particles stored in $21$  different snapshots to construct  our continuous past-light-cones up to $z=4$, building $27$ lens planes of the projected matter density distribution,  considering a  square sky  coverage of  five degrees on a side.  For  each  cosmological model  we construct $256$  different light-cone  realisations by  randomising the various  comoving  cosmological  boxes \citep{Giocoli_Baldi_Moscardini_2018,Peel2018}.

\subsection{Convergence maps}
\label{SEC::SIM::kappa}

The  \mapsim pipeline is composed of two algorithms. The first one -- called 
i-\mapsim -- constructs  lensing planes  from the different simulation  snapshots, saving  for each  plane $l$ and on each pixel  with coordinate  indices  $(i,j)$ the  particle surface  mass density $\Sigma$
\begin{equation}
\Sigma_l(i,j) = \dfrac{\sum_k m_k}{A_{l}}\,.
\end{equation}
$A_{l}$ represents the comoving pixel area of the lens plane $l$ and $\sum_k  m_k$ the sum over  all particle masses associated  with the given pixel. The second algorithm named ray-\mapsim projects the matter density distribution along the line-of-sight by weighing the  lens planes with the  lensing kernel in the Born  approximation regime \citep{Bartelmann2001,schaefer12,Petri2016,giocoli16a,petri17,giocoli17,giocoli18a,  castro17}. From $\Sigma_l$  we can derive  the convergence $\kappa$  at a given source redshift $z_s$ as
\begin{equation}
\kappa = \sum_l \dfrac{\Sigma_l}{\Sigma_{\rm{crit},l,s}}\,,
\end{equation}
where $l$ varies over the different lens planes with the lens redshift $z_l$ smaller  than $z_s$ and $\Sigma_{\rm{crit},l,s}$  represents the critical  surface density  at  the  lens plane  $z_l$  for sources  at redshift $z_s$
\begin{equation}
\Sigma_{\rm{crit},l,s} \equiv \frac{c^2}{4 \pi G} \frac{D_l}{D_s D_{ls}}.
\end{equation}
Here $c$ is the speed of  light, $G$ is Newton's constant and $D_l$, $D_s$ and  $D_{ls}$ are the angular  diameter distances between observer-lens, observer-source and source-lens, respectively. The final $\kappa$ maps cover the 25 square degree field-of-view with $2048^2$ pixels, resulting in a map resolution of $\sim 8.8$ arc-seconds.

\section{Methodology}
\label{SEC::METHOD}
A variety of machine learning techniques is applied to the \Dp~convergence maps. It was shown by \citet{Peel2018} that summary statistics up to second-order do not reliably separate such mass maps. Higher-order statistics, especially peak counts \citep[e.g.][]{Peel2017,Shan2018,Martinet2018,Lin2018}, do a better job but still leave room for improvement when distinguishing between a large number of models and in the presence of noise. Most commonly used methods to characterise observational data are naturally based on physical models. In the following we present an agnostic approach, which also applies techniques and algorithms found in the fields of computer science and specifically digital image processing. 

We distinguish two subsequent steps in the process of mass map classification. The first is to find a feature extraction function $\Theta$, which takes a high-dimensional data vector $x$ as input and finds a general, dimensional reduced representation of it in the form of a feature vector $F$
\begin{equation}
\label{EQU::METHOD::feature}
\Theta(x;\wf) = F.
\end{equation}
The feature extraction function can have several parameters which are stored in the feature weight vector $\wf$.
In order to arrange the data vector $x$ in a meaningful way, we introduce an index notation $x_{ijc}$. The first two indices reflect a spatial ordering of the 2D data along the coordinate axes. This means that all elements with $i=1$ are located in the first row of the pixelised image and all elements with e.g. $j = 10$ are located in the tenth column of the image. This notation also includes 1D data, ordered or not, by setting $i=1 ~\forall j,c$. The third index $c$ --commonly dubbed as a channel-- allows us to collect multiple aspects of the same entity represented by $x$. For the example of an RGB-image, $c=1$ would be the red channel of the image, $c=2$ the green and $c=3$ the blue channel. Finally, we define the shape of a data vector with a bracket notation. The shape of our input convergence maps is $\#x=(2048,2048,6)$ since we have $2048\times 2048$ pixel maps with convergence values $\kappa$ at six different source redshift channels and where in the above we have introduced the shape operator $\#$ which returns the shape of a data vector. 

The second step classifies $F$ into a set of target classes. The classification function $\zeta$, which can again depend on a set of parameters $\wc$, should not only output a single class prediction, but rather a prediction vector $P$ of shape $\#P=(1,1,n)$ with probabilities to belong to each of $n$ target classes
\begin{equation}
\label{EQU::METHOD::prediction}
\zeta(F,\wc) = P.
\end{equation}
It must hold that $P_{n} \in [0,1]$, $\sum P_{n}=1$ and in our case $n\in(1,...,9)$. 

In the following we explore different choices for the feature extraction and classification functions and find ways to optimise their parameters to achieve an optimal classification. We do so with the help of training sets, which are data vector--label pairs $(x,y^{l})$, meaning mass maps for which we a-priori know the underlying cosmological model. Specifically, the label $y^{l}$ is an indicator function for the class $l \in (1,...,9)$ with elements $y^{l}_{k}$ for which 
\begin{equation}
\label{EQU::METHOD::label}
y^{l}_{k} =  
\begin{cases}
1 \quad \text{if} \quad k =l\\
0 \quad \text{else.}
\end{cases}
\end{equation}
 
\subsection{Definition of data sets}
\label{SEC::METHOD::datasets}
\begin{figure*}
\includegraphics[width=.45\textwidth]{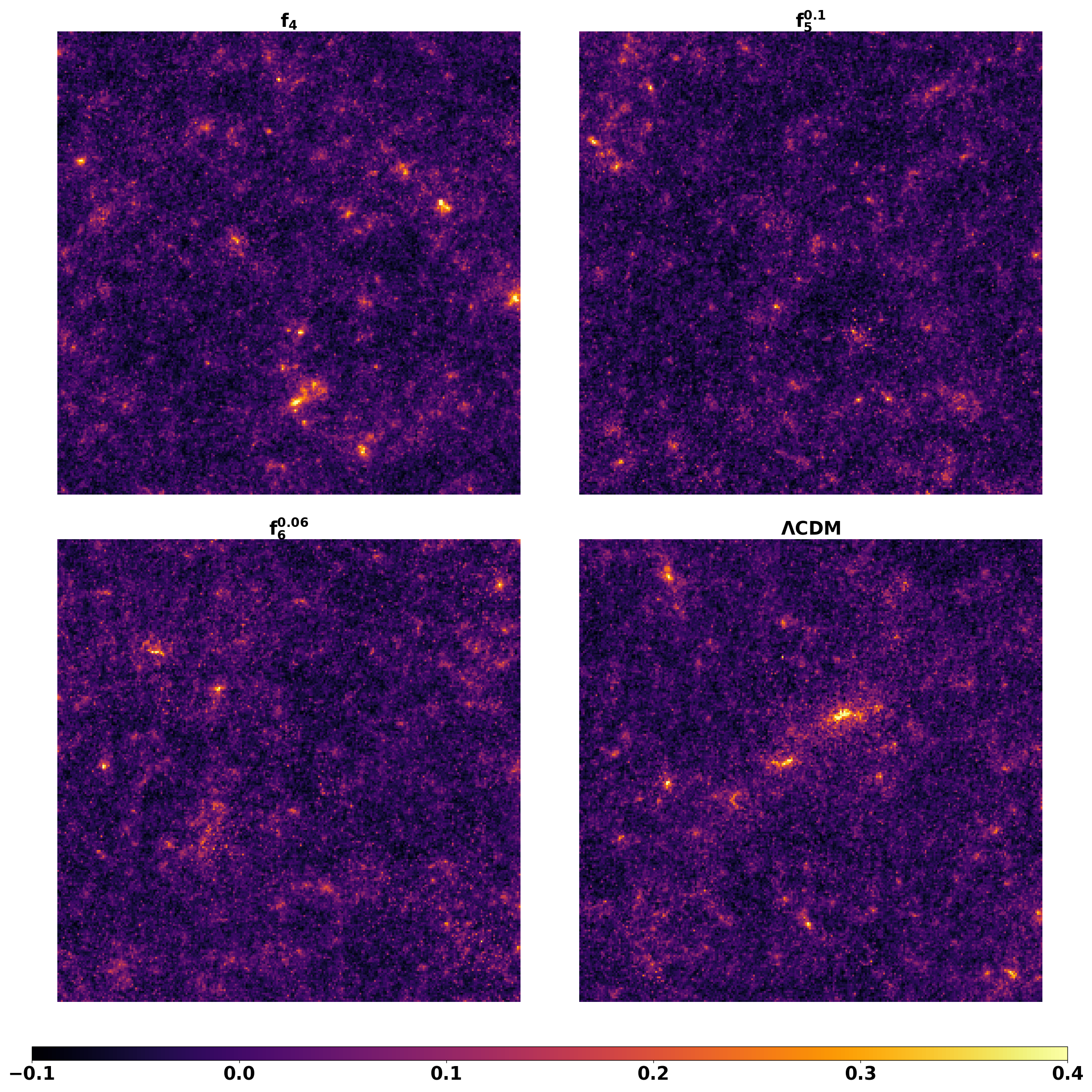}
\includegraphics[width=.45\textwidth]{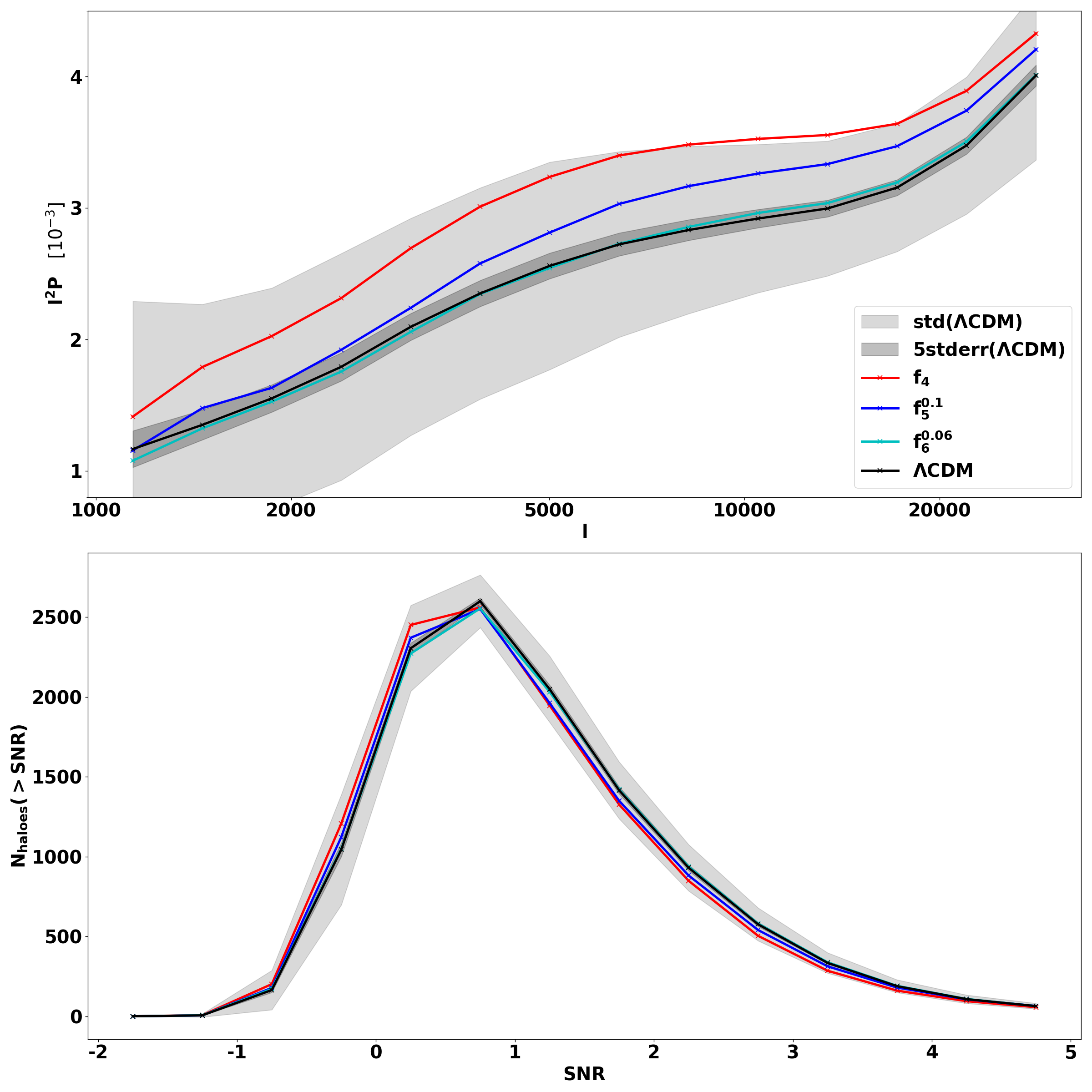}
\caption{The \textit{left panel} shows randomly chosen 
convergence maps from the test set in four specifically chosen models of 
structure formation. Those models span the range of observational 
degeneracy from the \lcdm reference, with $f_{4}$ being the most 
distinct, $f_{6}^{0.06}$ the most similar and $f_{5}^{0.1}$ the middle ground 
between the extremes. The \textit{right panel} shows, for the same models,  the 
average power spectrum over all maps in the test set on the \textit{top} and the 
peak counts as a function of signal-to-noise on the \textit{bottom}. The 
coloured lines indicate the results for the four different models, while the 
grey shaded areas indicate typical variations within the test set sample for 
the case of $\Lambda$CDM. The source redshift in all cases is 
$z_{\textrm{s}}=1.0$.}
\label{FIG::METHOD::data}
\end{figure*}

Our full data set consists of 256 convergence maps of shape (2048,2048,6) for 
each of the nine cosmological  models. We split each map further into 64 smaller 
patches to define our main data vectors with $\#x=(256,256,6)$. 75\% of those 
maps (12289) are used as a training set in order to optimise the parameters of 
our models. We use 15\% of the maps (2457)  as a validation set where the 
correct labels $y$ are known to us, but not to the optimisation algorithm. 
Performing a classification on the validation data serves as quality control and 
helps us to decide if an optimisation is successful and when to stop it. Another 
10\% of the maps (1638) are used as a test set, where the labels are not known 
to us a-priori and to which our trained and validated algorithms are applied to 
blindly. The success rates on those test sets will be the main result of this 
work. We provide examples of the actual data in the left panel 
of figure~\ref{FIG::METHOD::data}, which shows example convergence maps, chosen 
at random from the test set, for four instructive models. This includes the 
\lcdm 
reference, $f_{4}$ which deviates most from \lcdm, 
$f_{6}^{0.06}$ which is observationally most degenerate with \lcdm and a sample 
map of $f_{5}^{0.1}$ which is between the two extremes. The source redshift for 
all the maps shown is $z_{\textrm{s}}=1.0$.

\subsection{Mass map feature extraction}
\label{SEC::METHOD::features}
Two important subclasses for $\Theta$ are possible. In the first, the parameters 
$\wf$ are free and can be optimised during a training phase. In the second, they 
are fixed. We want to highlight that we do not perform any initial 
transformations of the data, which have proven to be useful for the analysis of 
lensing mass maps. It was shown in e.g. \citet{Peel2018} that an aperture mass 
transformation \citep{Schneider1996,Schneider1998} can largely improve the 
discrimination power of certain statistics, but we want to stay as general and 
agnostic as possible at this stage and use the raw pixel data of the convergence 
maps as the initial data vector.  

\subsubsection{Standard mass map descriptors}
\label{SEC::METHOD::features::standard}
Examples of fixed feature extraction are the mass map descriptors which are commonly used in the cosmological community to describe convergence or shear catalogues. For the purposes of this work, such descriptors serve as the reference for other techniques that we apply.
We combine a number of mass map features, which we extract with the 
\texttt{LensTools} 
package\footnote{\href{https://github.com/apetri/LensTools}{
https://github.com/apetri/LensTools}} by \citet{Petri2016} into a feature vector 
of shape (1,1,99). The first four entries in this vector are the mean, variance, 
skewness and kurtosis of the convergence maps. This is followed by eleven 
percentiles between the 0th-percentile (the minimum) and the 100th-percentile 
(the maximum) in steps of ten percent. The normalised histogram (PDF) of the 
convergence values in each map is sorted with 14 bins and the value for each bin 
is appended to the feature vector. Next, we calculate the power spectrum in 14 
logarithmically spaced bins between $l = 1000$ and $l = 32000$, which cover the 
angular size and resolution of our mass maps. Finally, we use the standard 
deviation of each map to define 14 signal-to-noise bins between -2 and 5. For 
each such bin we calculate the peak counts, as well as the first three Minkowski 
functionals \citep[e.g.][and references therein]{Kratochvil2012,Petri2015}, 
which concludes our collection of 99 features.The right panel 
of figure \ref{FIG::METHOD::data} shows
examples for the variation between models for such classical features. 
Presented there is the average 
power spectrum and peak count for all maps in the test set and for the four 
instructive models we chose in section \ref{SEC::METHOD::datasets} for data 
visualisation purposes.   

\subsubsection{Classical computer vision}
\label{SEC::METHOD::features::cv}
We know from \citet{Peel2018} that at least some of the standard descriptors above are not optimally suited for the task at hand and it is, at this point, not entirely obvious how to define better ones. This is why we now aim to derive as many fixed features as possible. The publicly available \wnd~algorithm \citep{Shamir2008,Orlov2008,Shamir2010} was designed for the classification of microscopy images and derives a particularly large feature vector of shape (1,1,2919). This includes most of the common statistics and descriptors known to digital image processing. Many features are thereby not only calculated from the raw image, but from some of its alternative representations like the Fourier, Wavelet, Chebyshev or Edge transformation. Moreover, some features are also extracted from transformations of transformations. While we did state earlier that we do not want to vet our data with transformations, we want to point out that the listed transformations are by no means inspired by the mechanisms of lensing or structure formation. We refer the interested reader to \citet{Orlov2008} for the full description of the algorithm and the description of the full feature vector, but we do provide a short summary in appendix~\ref{SEC::APP1} and a compact overview in table \ref{TAB::APP1::features}. 

\subsubsection{Convolutional neural networks}
\label{SEC::METHOD::features::cnn}
As the class of feature extraction functions which are able to change their shape during the training process we chose multi-layered
neural networks \citep{Lecun2015,Goodfellow2016}. The input data vector $x$ is manipulated and eventually reduced in dimension by a long --deep-- chain of simple layers $\theta$, which implement a specific mathematical operation. The output of one layer, becomes the input of the following layer and contains its own set of parameters $w_{i}$. The set of all layer parameters becomes the feature parameter vector $\wf$. 
\begin{align}
\label{EQU::METHOD::cnn}
\Theta(x,\wf)&=\theta_{n}\circ\theta_{n-1}\circ...\circ\theta_{1} \\
\theta_{i}\circ\theta_{i-1}(\cdot) &\equiv \theta_{i}\left(\theta_{i-1}(\cdot,w_{i-1}),w_{i}\right) \\
\wf &\equiv \left\{w_{i}\right\}^{n}_{i=1}
\end{align}
Deep neural networks source their performance from the sheer number of layers they are comprised of and have gained much popularity in recent years. This is mainly due to the advancements in numerical performance by e.g.~exploiting many-core architectures\footnote{General Purpose Graphics Processing Units (GPGPU) are a popular example of a many-core architecture.}. This allows for the construction of particularly deep and complex networks with hundreds of millions of parameters. The functional forms of the layers that are used in a deep neural network depend on the problem at hand. For image classification, convolutional neural networks (CNN) have proven to be particularly useful \citep{Krizhevsky2012,Simonyan2014,Szegedy2014,He2015,Lin2017} and hence we chose this class of models for our purposes. The main functionality of a CNN is provided by a convolutional layer $\mathrm{Conv}(n,m,\Delta i,\Delta j,p,C)$ which applies a number of $C$ convolutions with kernel size $(n,m)$ to a 2D input vector $I_{ijc}$ with $\#I_{ijc}=(X,Y,l)$. The stride parameters $\Delta i$ and $\Delta j$ allow one to implement dimensional reduction and the parameter $p$ controls if the input data is padded ($p=v$) or unaltered ($p=s$). We provide a thorough mathematical definition of all deep neural network layers  used in this work, including the convolutional one, in appendix \ref{SEC::APP2::layers}.

Convolution layers are often followed by pooling layers for dimensional reduction. We implement average pooling layers $\mathrm{AvgPool}(n,m,\Delta i,\Delta j,p)$ which average entries of the 2D data vector within a window of size $(n,m)$, apply a stride defined by  $\Delta i$ and $\Delta j$ and follow the same padding scheme that was introduced earlier. Maximum pooling layers $\mathrm{MaxPool}(n,m,\Delta i,\Delta j,p)$ work in a similar manner but instead of the average they return the maximum within a given window. Both pooling layers exist also as global versions, indicated by GlobalMaxPool and GlobalAvgPool, where all entries per channel are considered for either the maximum or averaging operation.    

Up to this point, we only allowed for layers to be placed strictly sequential. In order to implement a horizontal layout, we connect several layers to the same input and combine their results $I_{ijc_{1}},...,I_{ijc_{n}}$ with the help of a concatenation layer $\mathrm{Concatenate}(I_{ijc_{1}},...,I_{ijc_{n}})$.
This concept of performing not only one operation at a given depth of the network but several has proven very successful for image classification as e.g. shown in \citet{Szegedy2014}, who dubbed such horizontal layers as Inception modules. 

The output of a layer can be followed by a non-linear activation function. For convolution and pooling layers we mainly deploy  rectangular linear units (ReLU) and we give the full detail about the activation functions used in this work in appendix \ref{SEC::APP2::acti}.
To avoid the network from overfitting, so-called dropout layers are introduced as a regularisation. In there, a given percentage of the elements of an input vector are chosen at random and are subsequently discarded from the output \citep{Srivastava2014}.
Finally, to compensate for fluctuations in the amplitudes of input vectors at different network depth, \citet{Ioffe2015} introduced the concept of batch normalisation which we also use after each convolutional layer. The output of the last layer in the CNN, the feature vector $F$, is used for classification in a final section of the network which is commonly referred to as top. The concrete architecture of the CNN that we use in this work is provided in section \ref{SEC::RESULTS::cnn} and appendix \ref{SEC::APP2::arch}.

\subsection{Feature-based classification}
\label{SEC::METHOD::class}
We now turn our attention to the classification function $\zeta(F;\wc)$. We investigate two different approaches to classification. The first one is a nearest-neighbour-search scheme based on distances in feature space. The other approach, based again on a class of neural networks, uses regression through a training set to find the optimal mapping between features and labels. 

\subsubsection{Feature space distances}
\label{SEC::METHOD::class::dist}
In the following we denote with $T$ all those feature vectors that belong to a sample from the training set and with  $T^{n}$ the subset which belongs only to class $n$ of the training set.
We calculate a Fisher discriminant \citep[e.g.][]{Bishop2006} to find suited classification weights $\wc$ for each individual feature $T_{i}$.   
\begin{equation}
\label{EQU::METHOD::fisher}
(\wc)_{i} = \frac{\sum\limits_{n=1}^{N}\left(\left<T_{i}\right>-\left<T_{i}^{n} \right>\right)^{2}}{\sum\limits_{n=1}^{N}(\sigma_{T_{i}}^{n})^{2}}\frac{N}{N-1}
\end{equation}
Here $N$ is the total number of classes and $(\sigma_{T_{i}}^{n})^{2}$ is the variance of the feature $i$ within class $n$.

Once we found the weights $\wc$ we can define a weighted nearest-neighbour distance (WNN) of any feature vector $F$ to all the classes $n$ in the training set. 
\begin{equation}
\label{EQU::METHOD::wnn}
d_{\mathrm{WNN}}^{n} = \min\limits_{T\in T^{n}} \sum\limits_{i=1}^{M}
(\wc)_{i}\left( F_{i} -T_{i} \right)^{2},
\end{equation}
where $M = |F|$ is the length of the feature vector. The problem with this WNN distance is the fact that it is based only on a single element in the training set, the one that minimises the sum in equation \ref{EQU::METHOD::wnn}. To remedy this, \citet{Orlov2006} introduced a weighted neighbour distance (WND), which takes into account the distance to all elements in the training set, but largely penalises large distances through the free parameter $b$
\begin{equation}
\label{EQU::METHOD::wnd}
d_{\mathrm{WND}}^{n} = \frac{\sum\limits_{T\in
T_{c}}\left[\sum\limits_{i=1}^{M} (w_{c})_{i}\left( F_{i} -T_{i}
\right)^{2}\right]^{b}}{|T^{c}|}.
\end{equation}
\citet{Orlov2006} found that the results do not strongly depend on $b$ once $b>2$ and that $b=5$ is a generally good, numerically stable, choice.
The final step in order to make predictions $P$ is to define a similarity using a distance of choice, e.g. WNN or WND,  and by normalising appropriately
\begin{equation}
\label{EQU::METHOD::sim}
P_{n} = \left(d_{n}\sum\limits_{i=1}^{N}(d_{i})^{-1} \right)^{-1}.
\end{equation}

\subsubsection{Fully connected neural networks}
\label{SEC::METHOD::class::nn}
A different approach to the classification task is another form of neural network (equation \ref{EQU::METHOD::cnn}). The main layer in such a neural network is a fully connected --sometimes called affine--layer FC(n), which implements a linear mapping between the input vector of length $m$ and the output vector of length $n$ using a matrix of $nm$ free parameters and an additional bias parameter (see appendix \ref{SEC::APP2::layers}). 

Such layers are again chained together and the last layer produces an output vector of the same length as the number of classes $N$. As before, in between those layers one may use dropout, activation and normalisation layers. 
The top of the network is followed by a specific activation function called a softmax (see appendix \ref{SEC::APP2::acti})
which provides the desired predictions $P_{n}$.

Since the optimal weights $\wc$ are found by a regression, we define a loss function $L$, which in the case of this classification problem is a categorical cross entropy
\begin{equation}
\label{EQU::METHOD::crossentropy}
L(x;\wc) = -\sum\limits_{n=1}^{N}y_{n}\log{P_{n}(x;\wc)}.
\end{equation}
$y$ are the labels for the elements in the training data $x$ and $P_{n}(\wc)$ their class predictions given a current set of parameters $\wc$. In order to minimise the loss, while continuously looping over the training data, we use a specific implementation of stochastic gradient descent called \texttt{ADAM} \citep{Diederik2014}. The gradients of our model $\frac{d\zeta}{d\wc}$ are thereby calculated via a back-propagation algorithm \citep{Rumelhart1986}. We end this description of our methodology by noting that for a full feature extraction and classification chain $P=\zeta\left(\Theta(x;\wf);\wc\right)$, with a CNN as $\Theta$ and a neural network as classifier $\zeta$, the classification and feature extraction weights can be optimised at the same time.  

\subsection{Numerical setup}
\label{SEC::METHOD::setup}
As mentioned earlier we use the \texttt{Python} package \texttt{LensTools}\footnote{\href{https://github.com/apetri/LensTools}{https://github.com/apetri/LensTools}} \citep{Petri2016} for the extraction  of the standard map descriptors from section \ref{SEC::METHOD::features::standard}. For the computer vision fixed features from section \ref{SEC::METHOD::features::cv} we slightly adapted the publicly available version of \texttt{wnd-charm}\footnote{\href{https://github.com/wnd-charm/wnd-charm}{https://github.com/wnd-charm/wnd-charm}}. We altered the \texttt{C++} version of the feature extraction algorithm to accept \texttt{FITS} files \citep{Hanisch2001} as an input image container with pixel values as double precision floating-point numbers. We then use the feature output files of \texttt{wnd-charm} as an input for our own distance-based classification pipeline written in \texttt{Python}. We make these routines publicly available in this repository\footnote{\href{https://bitbucket.org/jmerten82/mydnn}{https://bitbucket.org/jmerten82/mydnn}}. All deep learning elements of our analysis stack use the widely used \texttt{tensorflow}\footnote{\href{https://www.tensorflow.org/}{https://www.tensorflow.org/}} framework, which uses NVIDIA's cuDNN \citep{Chetlur2014} library to carry out tensor operations on GPUs. We pair a \texttt{tensorflow} backend with the high-level deep learning \texttt{Python} interface \texttt{keras}\footnote{\href{https://keras.io/}{https://keras.io/}} as a frontend. The network training was carried out on two NVIDIA Titan Xp GPUs. All convergence maps and \texttt{Jupyter}\footnote{\href{http://jupyter.org/}{http://jupyter.org/}} notebooks used to produce the results in this work are either linked to or publicly available in the aforementioned repository. In there, we refer the reader to the 'reproducible\_science' folder. 


\section{Results}
\label{SEC::RESULTS}
Section \ref{SEC::METHOD} introduced a number of methods to perform mass map characterisation and classification. We now present the results obtained by applying those methods and provide details on their training process with the help of the validation sets. For the most successful method we investigate the dependence of the results on the convergence map source redshift and we end this section with a closer look at the most relevant features which are extracted by the different methods. If not stated otherwise, the results in this section are based on training, validation and test set maps at a source redshift $z_{\mathrm{s}}=1.0$.  

\subsection{Classification based on feature distance}
For the case of the distance-based classifier from section 
\ref{SEC::METHOD::class::dist}, the training process is just the derivation of 
the Fisher weights shown in equation \ref{EQU::METHOD::fisher}. We calculate 
them using the training set and present the 20 top-ranked features for the 
classical descriptors in table \ref{TAB::RESULTS::classicweights} and for the 
\wnd~features in table \ref{TAB::RESULTS::weights}. For the first case, we see 
quite a mix of features in the top, with the power spectrum and peak counts 
being the most important ones. This result is nicely 
confirmed by the right panel of figure \ref{FIG::METHOD::data}, which shows 
that the 
bins with $5000 < l < 15000$ of the power spectrum indeed show a clear 
separation between the more degenerate models. For the \wnd~features 
however, the ranking is completely dominated by Zernike coefficients on 
transformations of the image, with a few contributions of Haralick textures. One 
should keep in mind though that we extract a total of 2919 features, out of 
which 51 have a weight $>0.1$, 868 have a weight $>0.01$ and only 193 features 
have a vanishing weight. It is the combination of all the non-zero weights which 
will lead to the distance-based classification later on.

\begin{table}
\begin{tabular}{lccc}
Rank  & Name & Index & Weight \\
\hline
\hline
1 &Power spectrum &11 &0.106\\
2 &" &10 &0.104\\
3 &" &9 &0.092\\
4 &" &12 &0.084\\
5 &Peak counts &13 &0.083\\
6 &Power spectrum &8 &0.078\\
7 &Peak counts &12 &0.078\\
8 &Power spectrum &7&0.065\\
9 &Peak counts &5&0.064\\
10 &Skewness &--&0.059\\
11 &Peak counts&14&0.055\\
12 &Power spectrum&6 &0.051\\
13 &Percentile&100&0.051\\
14 &Minkowski functional 1&14 &0.049\\
15 &Percentile&0&0.049\\
16 &Power spectrum &13 &0.046\\
17 &Minkowski functional 2&14 &0.045\\
18 &Peak counts&11&0.044\\
19 &Power spectrum&5&0.042\\
20 &Kurtosis&--&0.042\\
\end{tabular}
\caption{The top-ranked classical mass map features according to their Fisher score (equation \ref{EQU::METHOD::fisher}). The meaning of each feature and the explanation of its index can be found in section \ref{SEC::METHOD::features::standard}.}
\label{TAB::RESULTS::classicweights}
\end{table}

\begin{table}
\begin{tabular}{lcccc}
Rank & Transform & Name & Index & Weight \\
\hline
\hline
1 &F &Zernike coefficients &20 &0.285\\
2 &F &" &42 &0.270\\
3 &F(W) &" &50 &0.255\\
4 &F(E) &" &52 &0.242\\
5 &F(W) &" &21 &0.236\\
6 &F(E) &" &39 &0.214\\
7 &F &" &12 &0.205\\
8 &F(W) &" &22 &0.204\\
9 &F(E) &" &37 &0.196\\
10 &F(W) &" &56 &0.183\\
11 &F(E) &" &5 &0.174\\
12 &F(E) &" &28 &0.170\\
13 &W&Haralick textures &5 &0.169\\
14 &F &Zernike coefficients &17 &0.166\\
15 &F(E) &" &34 &0.166\\
16 &F(E) &Haralick textures &0 &0.164\\
17 &F(E) &" &14 &0.161\\
18 &F(E) &Zernike coefficients &24 &0.159\\
19 &F&" &60 &0.154\\
20 &-- &Edge features &0 &0.152\\
\end{tabular}
\caption{The same as table \ref{TAB::RESULTS::classicweights} but for the \wnd~features. We refer the reader to appendix \ref{SEC::APP1} for the definition of each feature and the exact meaning of the feature index and the transform column.}
\label{TAB::RESULTS::weights}
\end{table}

For the 99 standard features we find a total classification success rate of 22\%, meaning that out of 14742 samples in the test set, only 3243 were classified correctly. For some specific classes the classification success rate is barely above the success rate for a random guess (11\%). The important \lcdm class for example shows a success rate of 13\%. The picture improves marginally when using the 2919 \wnd~features instead. The total classification success over all classes rises mildly to 25\%. While especially the three $f_{6}$ models still show success rates around or even below 11\%, at least some classes, including \lcdm, are now significantly above the 20\% level. We do not show more details\footnote{A full success rate analysis is provided in the repsository (\href{https://bitbucket.org/jmerten82/mydnn}{https://bitbucket.org/jmerten82/mydnn}) associated with this article.} on the distance-based classification since it is clear from those results already that this classification method does not qualify for a successful discrimination of our models. 

\subsection{Classification based on neural network}
\begin{figure}
\includegraphics[width=.45\textwidth]{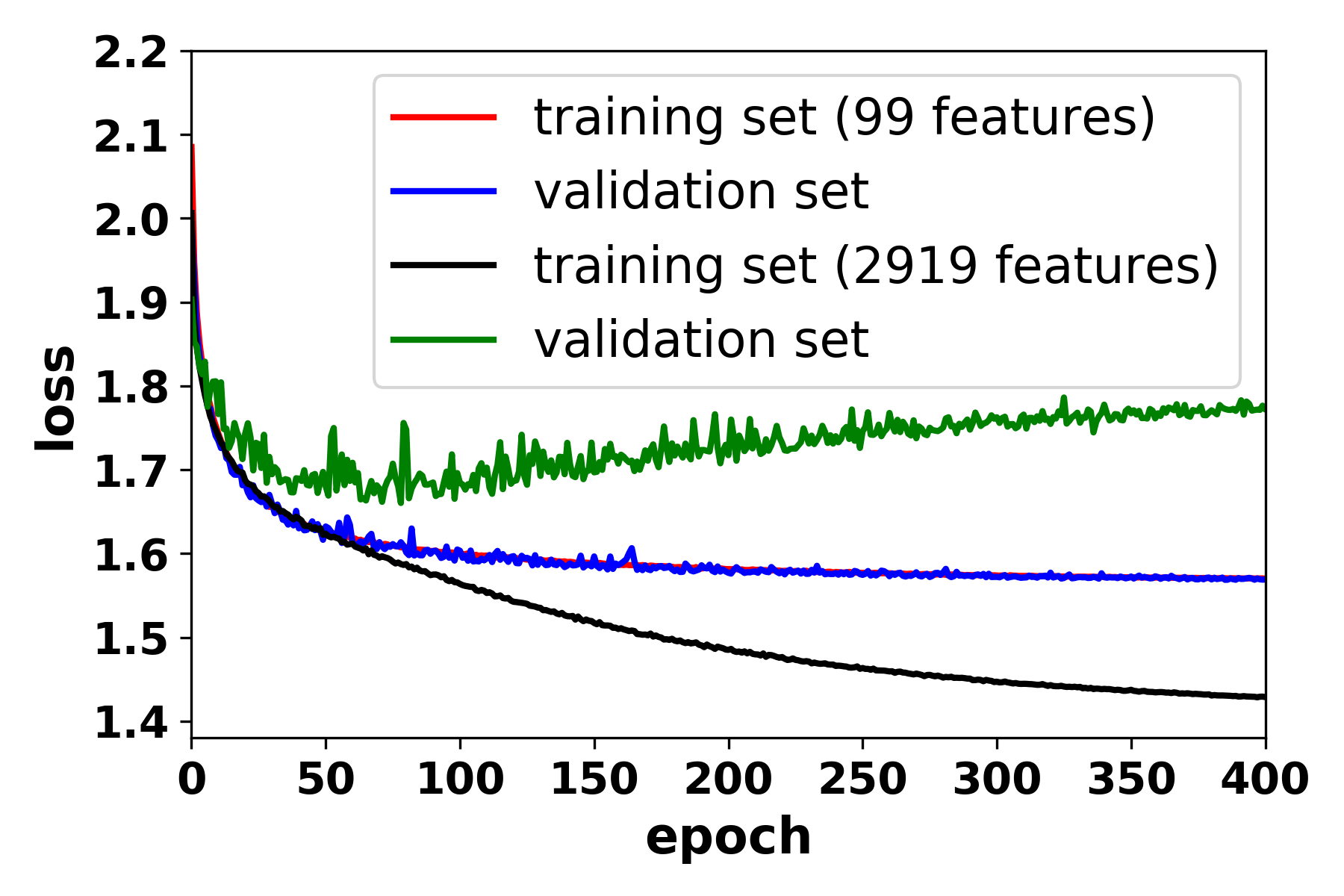}
\caption{The evolution of the loss as a function of epoch for the training of the neural network. Shown are both cases where either a smaller vector of standard is the input for the network, or a larger set of  \wnd~features.}
\label{FIG::RESULTS::nn_training}
\end{figure}

We now use the same set of fixed features but feed them into a fully-connected neural network for classification. For the case of the 99 standard features we show the very simple topology of the classification network in table \ref{TAB::RESULTS::classic_nn}. The same network is used to classify the 2919 \wnd~features but due the larger input vector, the number of free parameters is larger, which we indicate by a square bracket notation in the same table. The regression to find the optimal parameters of the main fully-connected layers is based on the training set. In total we train with 110601 feature vectors of shape (1,1,99) or (1,1,2919) and where one iteration over all those elements during the regression is commonly called an epoch. Gradient evaluations and corresponding changes to the network parameters are made after a subset of an epoch, usually called a batch. The batch size in this case was set to 128.  After each epoch, we evaluate the current performance of the network with the 22113 (2457 per class) feature vectors in the validation set. Figure \ref{FIG::RESULTS::nn_training} shows for both feature sets the evolution of the loss function for the training and validation data as a function of training epoch. For the larger feature vector, the validation loss starts to saturate around epoch 70 while the training loss keeps declining. This indicates that the networks starts to overfit, meaning that it learns training-set specific features which are of no use to characterise the validation set or any data unknown to the model. This is where we stop the training and save the model parameters which produced the smallest validation loss.

\begin{table}
\begin{tabular}{lccc}
Index&Layer&free parameters&Output shape\\
\hline
\hline
1 & Input & 0 & (1,1,99[2919]) \\
\hline
2 & FC(32) & 3200 [93440] & (1,1,32) \\
3 & leakyReLU(0.03) &0 & (1,1,32) \\
4 & FC(9)& +297 & (1,1,9) \\
5 & Softmax &0 & (1,1,9) \\
\hline
 & Ouput & =3497 [93737]  &9 
\end{tabular}
\caption{The sequence of layers used in the neural network to classify fixed mass map features. The output shape notation follows the convention introduced in section \ref{SEC::METHOD}. The description of all layers can be found in sections \ref{SEC::METHOD::features::cnn} and \ref{SEC::METHOD::class::nn}, their formal definition in appendix \ref{SEC::APP2::layers} and \ref{SEC::APP2::acti}. The numbers in square brackets refer to the case where the \wnd~feature vector is used instead of the smaller vector of classical features.}
\label{TAB::RESULTS::classic_nn}
\end{table}  

The neural network classification yields significantly better results compared to the classification based on feature-space distances. In the case of the 99 standard features the total classification rate rises to 39\% and to 35\% in the case of the \wnd~features. Most interestingly, the smaller vector of 99 classical features produces better results than the much larger feature vector provided by \wnd. Some of the most discriminative features from the computer vision method shown in table \ref{TAB::RESULTS::weights} are certainly describing the data well and should be used in future analyses; however, once the information from all standard descriptors such as the binned power spectrum, peak counts and Minkowski functionals are combined in an optimal way by a neural network, there is no advantage in using features that are inspired by computer vision only. Table \ref{TAB::RESULTS::nn_cmatrix} shows the classification success matrix for the standard features, where each row refers to a subset of the test data comprising only maps from that true class labelled by the first column. The first number in each block of four shows how many times the 1638 members of this subset have been sorted into the respective predicted class which is indicated by the label in the very first row. The second number is the percentage of predictions with respect to the total number of maps in the class. The third and fourth numbers are the mean and its standard error on the prediction probability for all maps in the subset given by the row and for the class predictions indicated by the label of the column. For an optimal classification, only the diagonal of this matrix (those fields typeset in boldface) would show non-zero values. 

\begin{table*}
\begin{tabular}{l|ccccccccc}
&$f_{4}$&$f_{4}^{0.3}$&$f_{5}$&$f_{5}^{0.15}$&$f_{5}^{0.1}$&$f_{6}$&$f_{6}^{0.06}$&$f_{6}^{0.1}$&$\Lambda$CDM\\
\hline
$f_{4}$& \textbf{958}&80&157&103&97&137&26&24&56\\
&\textbf{58\%}&5\%&10\%&6\%&6\%&8\%&2\%&1\%&3\%\\
&$\mathbf{0.376}$&$0.052$&$0.122$&$0.083$&$0.11$&$0.087$&$0.068$&$0.056$&$0.046$\\
&$\mathbf{\pm0.007}$&$\pm0.003$&$\pm0.003$&$\pm0.002$&$\pm0.002$&$\pm0.002$&$\pm0.002$&$\pm0.001$&$\pm0.002$\\ \hline
$f_{4}^{0.3}$&70& \textbf{1135}&1&72&10&18&31&116&185\\
&4\%&\textbf{69\%}&0\%&4\%&1\%&1\%&2\%&7\%&11\%\\
&$0.05$&$\mathbf{0.468}$&$0.006$&$0.052$&$0.018$&$0.058$&$0.091$&$0.12$&$0.137$\\
&$\pm0.002$&$\mathbf{\pm0.007}$&$\pm0.001$&$\pm0.002$&$\pm0.001$&$\pm0.001$&$\pm0.002$&$\pm0.002$&$\pm0.003$\\ \hline
$f_{5}$&232&6& \textbf{857}&158&275&84&10&5&11\\
&14\%&0\%&\textbf{52\%}&10\%&17\%&5\%&1\%&0\%&1\%\\
&$0.129$&$0.008$&$\mathbf{0.354}$&$0.129$&$0.246$&$0.055$&$0.036$&$0.026$&$0.018$\\
&$\pm0.004$&$\pm0.001$&$\mathbf{\pm0.005}$&$\pm0.003$&$\pm0.002$&$\pm0.002$&$\pm0.001$&$\pm0.001$&$\pm0.001$\\ \hline
$f_{5}^{0.15}$&149&90&173& \textbf{651}&254&120&38&52&111\\
&9\%&5\%&11\%&\textbf{40\%}&16\%&7\%&2\%&3\%&7\%\\
&$0.082$&$0.052$&$0.129$&$\mathbf{0.229}$&$0.182$&$0.087$&$0.086$&$0.08$&$0.073$\\
&$\pm0.003$&$\pm0.003$&$\pm0.003$&$\mathbf{\pm0.003}$&$\pm0.003$&$\pm0.002$&$\pm0.002$&$\pm0.002$&$\pm0.002$\\ \hline
$f_{5}^{0.1}$&223&22&485&333& \textbf{413}&98&18&14&32\\
&14\%&1\%&30\%&20\%&\textbf{25\%}&6\%&1\%&1\%&2\%\\
&$0.113$&$0.02$&$0.249$&$0.18$&$\mathbf{0.243}$&$0.067$&$0.054$&$0.043$&$0.031$\\
&$\pm0.003$&$\pm0.001$&$\pm0.005$&$\pm0.003$&$\mathbf{\pm0.003}$&$\pm0.002$&$\pm0.001$&$\pm0.001$&$\pm0.001$\\ \hline
$f_{6}$&193&103&44&181&65& \textbf{512}&126&171&243\\
&12\%&6\%&3\%&11\%&4\%&\textbf{31\%}&8\%&10\%&15\%\\
&$0.091$&$0.061$&$0.048$&$0.094$&$0.065$&$\mathbf{0.192}$&$0.17$&$0.155$&$0.123$\\
&$\pm0.003$&$\pm0.003$&$\pm0.002$&$\pm0.002$&$\pm0.002$&$\mathbf{\pm0.003}$&$\pm0.002$&$\pm0.002$&$\pm0.003$\\ \hline
$f_{6}^{0.06}$&128&194&30&163&39&377& \textbf{147}&257&303\\
&8\%&12\%&2\%&10\%&2\%&23\%&\textbf{9\%}&16\%&18\%\\
&$0.07$&$0.095$&$0.032$&$0.087$&$0.05$&$0.169$&$\mathbf{0.176}$&$0.176$&$0.146$\\
&$\pm0.003$&$\pm0.003$&$\pm0.002$&$\pm0.002$&$\pm0.002$&$\pm0.002$&$\mathbf{\pm0.002}$&$\pm0.002$&$\pm0.003$\\ \hline
$f_{6}^{0.1}$&71&262&18&164&39&279&109& \textbf{345}&351\\
&4\%&16\%&1\%&10\%&2\%&17\%&7\%&\textbf{21\%}&21\%\\
&$0.049$&$0.12$&$0.023$&$0.086$&$0.042$&$0.154$&$0.176$&$\mathbf{0.189}$&$0.162$\\
&$\pm0.002$&$\pm0.004$&$\pm0.001$&$\pm0.002$&$\pm0.002$&$\pm0.002$&$\pm0.002$&$\mathbf{\pm0.002}$&$\pm0.003$\\ \hline
$\Lambda$CDM&69&265&5&135&9&158&60&166& \textbf{771}\\
&4\%&16\%&0\%&8\%&1\%&10\%&4\%&10\%&\textbf{47\%}\\
&$0.043$&$0.14$&$0.014$&$0.074$&$0.028$&$0.12$&$0.146$&$0.164$&$\mathbf{0.271}$\\
&$\pm0.002$&$\pm0.004$&$\pm0.001$&$\pm0.002$&$\pm0.001$&$\pm0.002$&$\pm0.002$&$\pm0.002$&$\mathbf{\pm0.004}$\\ \hline
\end{tabular}
\caption{The classification success matrix for the neural-network-based classification of the classical features. Each row represents a different subset of the test data indicated by the first column. The first number in each block of four in a column is the number of samples in the subset that was assigned to the predicted class indicated by the column label on the top. The second number is the relative classification success rate for the subset. The third number is the mean of all predictions in the subset and the fourth number is its standard error. The success rates indicate that only the two $f_{4}$ models and to a lesser degree $f_{5}$, $f_{5}^{0.15}$ and \lcdm are well separated from the other models with correct classification rates of 40\% or above and false classification rates for other models of 17\% or less. $f_{5}^{0.1}$ and the two  $f_{6}$ models with non-vanishing neutrino mass are basically undistinguished from other models and $f_{6}$ (success rate 31\%) shows still a large degeneracy with \lcdm (15\% misclassification rate).}
\label{TAB::RESULTS::nn_cmatrix}
\end{table*}

\begin{figure}
\includegraphics[width=.5\textwidth]{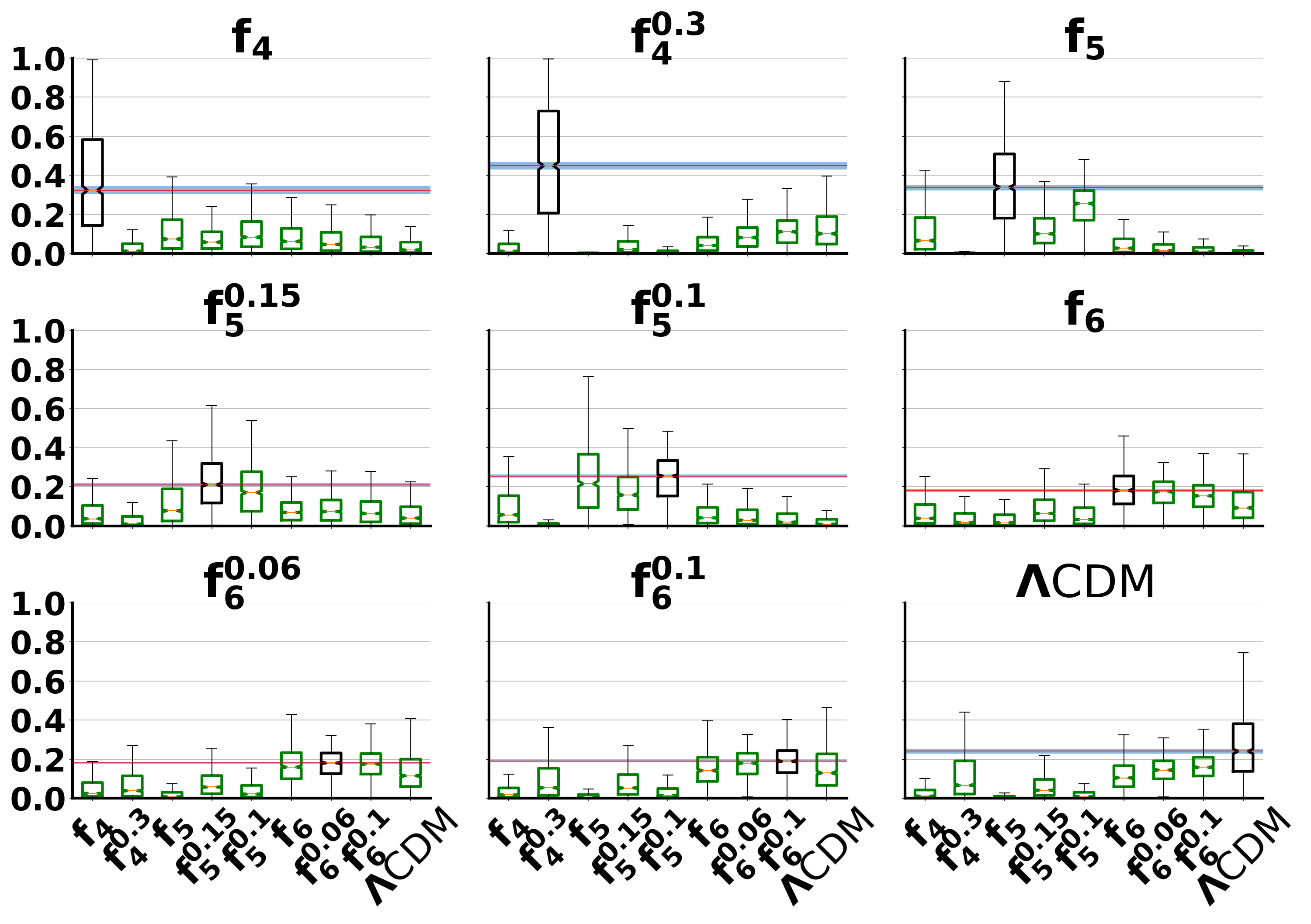}
\caption{The prediction statistics of the classical feature vector classified by a neural network. Each labelled panel represents all predictions for one true class of the test set. In every panel, each box summarises the statistics of the model predictions indicated by the bottom labels. The median and its bootstrap error for the correct prediction is shown by the red line with error band. For this method only the two $f_{4}$ models are clearly distinguished from the other models. Especially the $f_{5}$ and $f_{6}$ models remain largely degenerate within themselves, but also with $\Lambda$CDM.}
\label{FIG::RESULTS::nn_box}
\end{figure}

While table \ref{TAB::RESULTS::nn_cmatrix} gives a good indication of what to expect from a classification of single maps, only the mean and its standard error on the class predictions give an idea on how well the full ensemble of test set maps is classified. We therefore further evaluate the statistics of the prediction vectors for each true test set class. Figure \ref{FIG::RESULTS::nn_box} shows nine panels of box plots, each of which represents the statistics for one such subset. The black box in each panel represents the correct predictions, equivalent to the bold diagonal of table \ref{TAB::RESULTS::nn_cmatrix}. The horizontal line spanning each panel is the median for all the true class predictions and the error band shows the scatter of medians derived from 1000\footnote{This number is of course arbitrary but is close to the sample size and we also checked that the bootstrap-derived error does not depend significantly on the number of bootstraps.} bootstrap samples. The upper and lower end of each box show the 75th and 25th percentiles, respectively, and the whiskers show the outlier cleaned minimum and maximum value of the class predictions. Whenever a box which is not the true label is shown in green, it means that the median and its errors, indicated by the notches of each box, is lower than the one of the correct prediction box and does not overlap with its horizontal error band. If those criteria are not met, the respective box is shown in red.

When looking at the results in table \ref{TAB::RESULTS::nn_cmatrix} and figure \ref{FIG::RESULTS::nn_box}, the following observations catch the eye. Although the overall classification success rate is only 39\%, none of the classes is classified incorrectly as an ensemble. In the case of the two $f_{4}$ models we see a clear separation between the correct predictions from the other classes. This is confirmed by the classification matrix, which shows no substantial overlap ($> 11\%$) with any other model. This however changes for the three $f_{5}$ and the three $f_{6}$ models. Although the median for the correct predictions is the highest for all of the models\footnote{The mean is not in the case of $f_{5}^{0.1}$}, the degeneracies within the same model of gravity are strong in those cases as one can see from the basically equal heights of the centres of the boxes in figure \ref{FIG::RESULTS::nn_box} and from the classification matrix, which lists a large number of misclassifications up to 30\% in the case of $f_{5}^{0.1}$ misclassified as $f_{5}$. A lot more severe is the case of the three $f_{6}$ models. For them we find substantial overlap of up to 21\% with \lcdm. Even the predictions for \lcdm itself are not completely separate from the three $f_{6}$ models and $f_{4}^{0.3}$ with an overlap of up to 16\%.

\subsection{Convolutional neural network}
\label{SEC::RESULTS::cnn}
\begin{figure}
\includegraphics[width=.45\textwidth]{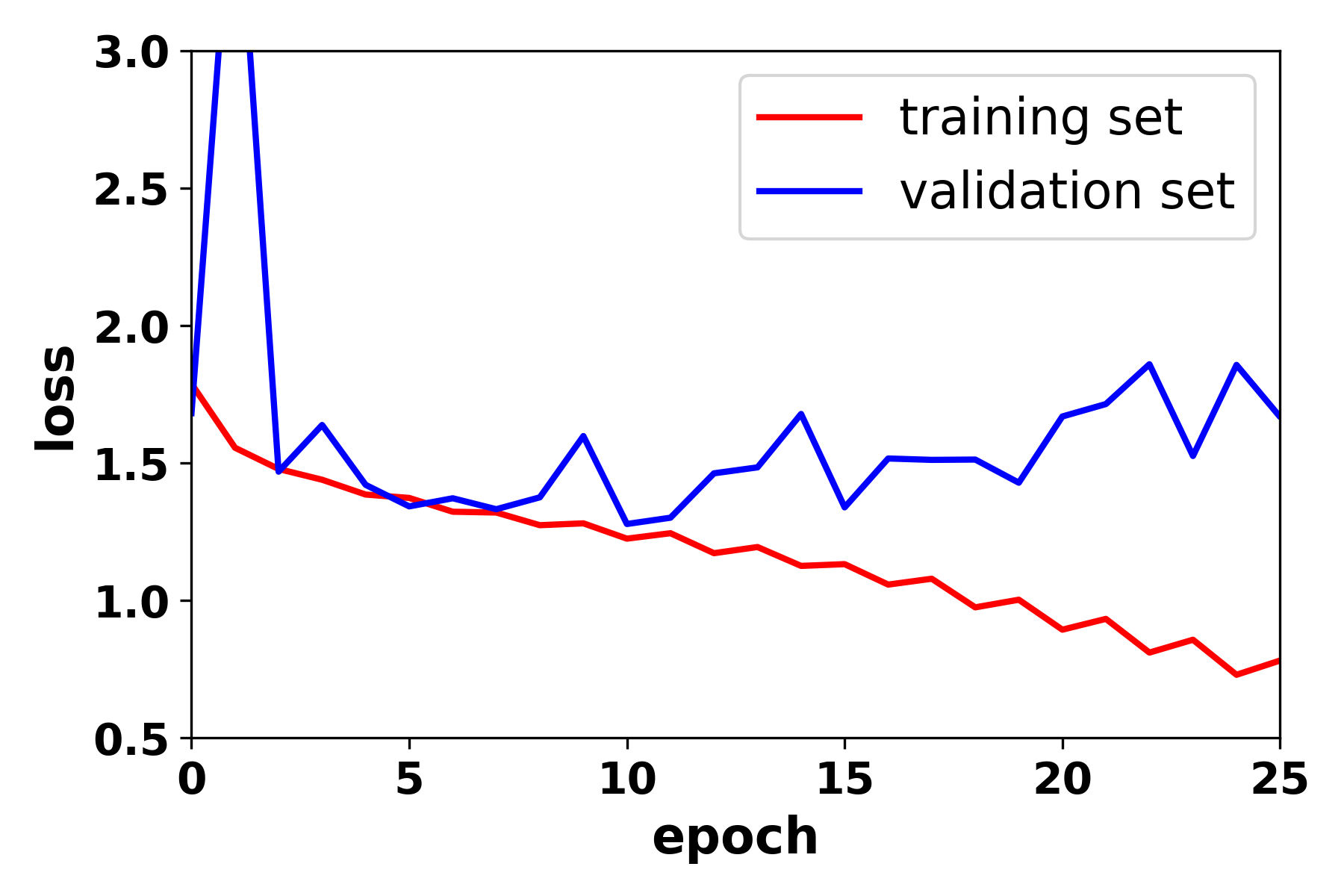}
\caption{The evolution of the loss as a function of epoch for the training of the CNN.}
\label{FIG::RESULTS::cnn_training}
\end{figure}

The convolutional neural network (CNN) extracts the characterising features directly from the pixel data of the training mass maps. We have experimented with a number of architectures, including classic topologies which implement a large number of $3\times3$ convolutions inspired by VGG-net \citep{Simonyan2014}, as well as architectures presented in \citet{Ravanbakhsh2017} and \citet{Gupta2018}. The model that worked best for our purposes is almost exclusively based on the Inception layers first presented in \citet{Szegedy2014}. Here we adopt one of its latest iterations, version 4 introduced in \citet{Szegedy2016}. The global linear structure of our CNN is shown in table \ref{TAB::RESULTS::cnn} and we describe in detail  the different elements of this network and their purpose in appendix \ref{SEC::APP2::arch}.
\begin{table}
\begin{tabular}{lccc}
&Layer&free parameters&Output shape\\
\hline
\hline
1 & Input & 0 & (256,256,1)\\
\hline
2 & Conv(3,3,2,2,v,32)$^{*}$ & 288 & (127,127,32)\\
3 & lReLU(0.03) &0 & (127,127,32) \\
4 & Conv(3,3,1,1,v,32)$^{*}$ & +9216 & (125,125,32)\\
5 & lReLU(0.03) &0 & (125,125,32) \\
6 & Conv(3,3,1,1,s,64)$^{*}$ & +9216 & (125,125,32)\\
7 & StemInception$^{*}$ (Fig.~\ref{FIG::RESULTS::stem}) & +555008 & (29,29,384) \\
8 &InceptionA$^{*}$ (Fig.~\ref{FIG::RESULTS::a}) & +316416 & (29,29,384) \\
9 &ReductionA$^{*}$ (Fig.~\ref{FIG::RESULTS::ra}) & +2304000 & (14,14,1024) \\
10 &InceptionB$^{*}$ (Fig.~\ref{FIG::RESULTS::b}) & +2931712 & (14,14,1024) \\
11 &ReductionB$^{*}$ (Fig.~\ref{FIG::RESULTS::rb})  & +2744320 & (6,6,1536) \\
12 &InceptionC$^{*}$ (Fig.~\ref{FIG::RESULTS::c})& +4546560 & (6,6,1536) \\
13 & GlobalAvgPool & 0 & (1,1,1536)\\
14 & Dropout(0.33) & 0 & (1,1,1536)\\
15 &FC(9) & +13833 & (1,1,9) \\
16 &Softmax & 0 & (1,1,9)\\
\hline
17 & Ouput & =13469865  &9 
\end{tabular}
\caption{The sequential structure of the CNN used in this work. All layers marked by $^{*}$ are batch normalised. More complicated Inception layers are shown in the respective figure.}
\label{TAB::RESULTS::cnn}
\end{table}
We visualise the evolution of the network's  loss during training in figure \ref{FIG::RESULTS::cnn_training}. 

\begin{figure}
\includegraphics[width=.5\textwidth]{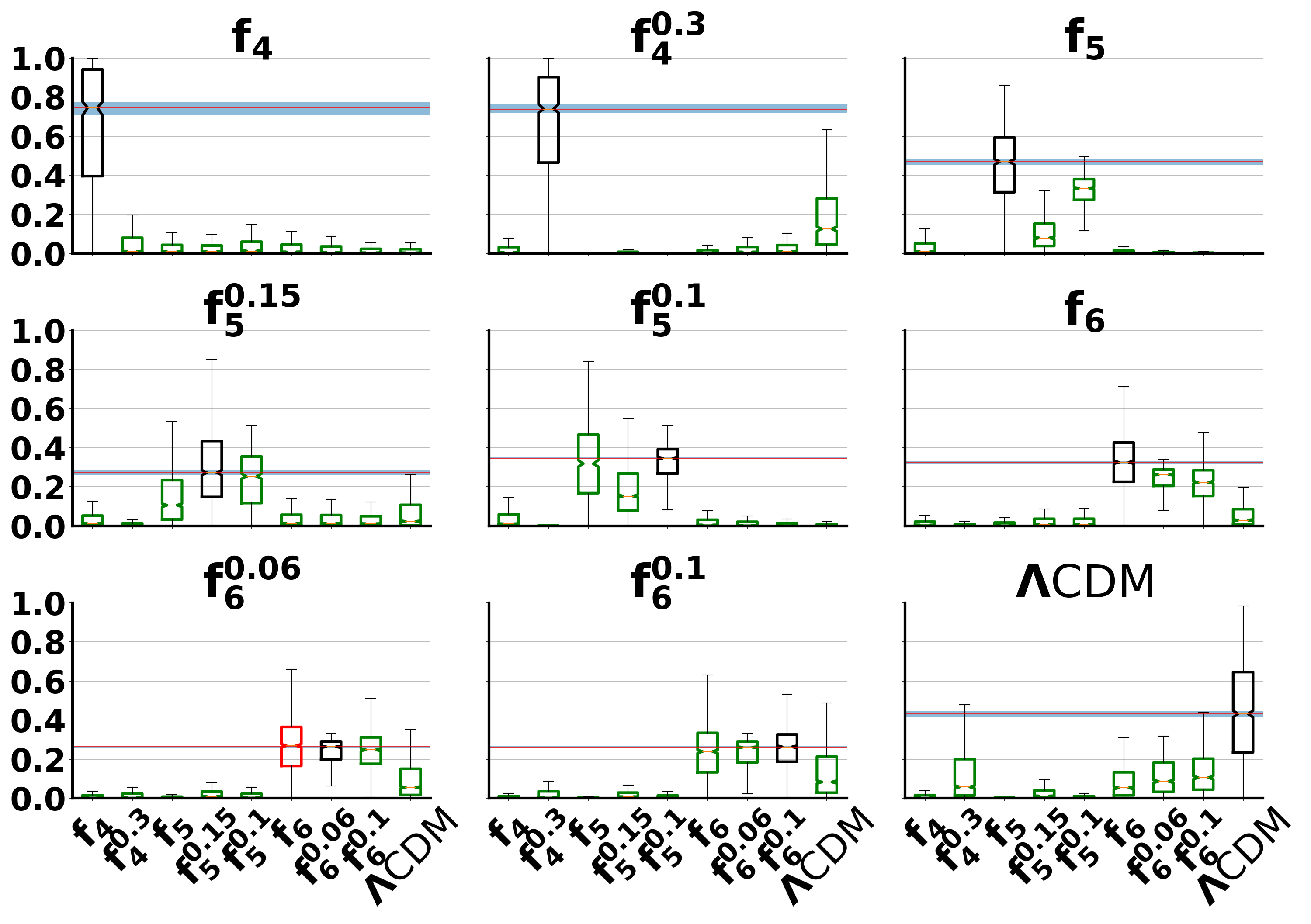}
\caption{Prediction statistics for the CNN at source redshift \zsone. The structure of the figure is the same as in figure \ref{FIG::RESULTS::nn_box}. The CNN discriminates more clearly between the models and both $f_{4}$ models, $f_{5}$ and \lcdm are now clearly distinguished. Problems remain for the different neutrino mass realisations within $f_{6}$ and $f_{5}$ gravity. $f_{6}^{0.06}$ is incorrectly classified as $f_{6}^{0.1}$.}
\label{FIG::RESULTS::cnn_box}
\end{figure}

The total classification success rate of the CNN is 52\% and its classification success matrix is shown in table \ref{TAB::RESULTS::cnn_cmatrix}.
\begin{table*}
\begin{tabular}{l|ccccccccc}
&$f_{4}$&$f_{4}^{0.3}$&$f_{5}$&$f_{5}^{0.15}$&$f_{5}^{0.1}$&$f_{6}$&$f_{6}^{0.06}$&$f_{6}^{0.1}$&$\Lambda$CDM\\
\hline
$f_{4}$& \textbf{1307}&116&36&21&31&88&12&5&22\\
&\textbf{80\%}&7\%&2\%&1\%&2\%&5\%&1\%&0\%&1\%\\
&$\mathbf{0.646}$&$0.085$&$0.045$&$0.035$&$0.05$&$0.049$&$0.036$&$0.025$&$0.028$\\
&$\mathbf{\pm0.008}$&$\pm0.004$&$\pm0.002$&$\pm0.002$&$\pm0.002$&$\pm0.002$&$\pm0.002$&$\pm0.001$&$\pm0.002$\\ \hline
$f_{4}^{0.3}$&51& \textbf{1298}&0&11&0&7&8&27&236\\
&3\%&\textbf{79\%}&0\%&1\%&0\%&0\%&0\%&2\%&14\%\\
&$0.046$&$\mathbf{0.658}$&$0.001$&$0.014$&$0.004$&$0.02$&$0.031$&$0.037$&$0.19$\\
&$\pm0.003$&$\mathbf{\pm0.007}$&$\pm0.0$&$\pm0.001$&$\pm0.0$&$\pm0.001$&$\pm0.001$&$\pm0.002$&$\pm0.005$\\ \hline
$f_{5}$&105&1& \textbf{1065}&90&320&48&2&1&6\\
&6\%&0\%&\textbf{65\%}&5\%&20\%&3\%&0\%&0\%&0\%\\
&$0.064$&$0.002$&$\mathbf{0.444}$&$0.114$&$0.316$&$0.028$&$0.016$&$0.011$&$0.006$\\
&$\pm0.003$&$\pm0.0$&$\mathbf{\pm0.005}$&$\pm0.003$&$\pm0.002$&$\pm0.002$&$\pm0.001$&$\pm0.001$&$\pm0.001$\\ \hline
$f_{5}^{0.15}$&103&37&161& \textbf{721}&347&78&14&31&146\\
&6\%&2\%&10\%&\textbf{44\%}&21\%&5\%&1\%&2\%&9\%\\
&$0.059$&$0.031$&$0.153$&$\mathbf{0.3}$&$0.235$&$0.048$&$0.044$&$0.041$&$0.088$\\
&$\pm0.003$&$\pm0.002$&$\pm0.004$&$\mathbf{\pm0.005}$&$\pm0.003$&$\pm0.002$&$\pm0.002$&$\pm0.002$&$\pm0.004$\\ \hline
$f_{5}^{0.1}$&122&5&624&271& \textbf{514}&70&5&7&20\\
&7\%&0\%&38\%&17\%&\textbf{31\%}&4\%&0\%&0\%&1\%\\
&$0.071$&$0.007$&$0.323$&$0.187$&$\mathbf{0.315}$&$0.036$&$0.025$&$0.019$&$0.018$\\
&$\pm0.004$&$\pm0.001$&$\pm0.005$&$\pm0.004$&$\mathbf{\pm0.003}$&$\pm0.002$&$\pm0.001$&$\pm0.001$&$\pm0.001$\\ \hline
$f_{6}$&51&27&11&42&44& \textbf{968}&74&307&114\\
&3\%&2\%&1\%&3\%&3\%&\textbf{59\%}&5\%&19\%&7\%\\
&$0.035$&$0.022$&$0.023$&$0.034$&$0.036$&$\mathbf{0.326}$&$0.235$&$0.218$&$0.072$\\
&$\pm0.002$&$\pm0.002$&$\pm0.001$&$\pm0.002$&$\pm0.002$&$\mathbf{\pm0.004}$&$\pm0.002$&$\pm0.002$&$\pm0.003$\\ \hline
$f_{6}^{0.06}$&36&46&11&40&35&713& \textbf{95}&458&204\\
&2\%&3\%&1\%&2\%&2\%&44\%&\textbf{6\%}&28\%&12\%\\
&$0.026$&$0.036$&$0.017$&$0.034$&$0.029$&$0.271$&$\mathbf{0.235}$&$0.24$&$0.112$\\
&$\pm0.002$&$\pm0.002$&$\pm0.001$&$\pm0.002$&$\pm0.002$&$\pm0.003$&$\mathbf{\pm0.002}$&$\pm0.002$&$\pm0.004$\\ \hline
$f_{6}^{0.1}$&20&79&5&35&17&558&64& \textbf{565}&295\\
&1\%&5\%&0\%&2\%&1\%&34\%&4\%&\textbf{34\%}&18\%\\
&$0.018$&$0.05$&$0.01$&$0.03$&$0.02$&$0.24$&$0.23$&$\mathbf{0.253}$&$0.149$\\
&$\pm0.001$&$\pm0.003$&$\pm0.001$&$\pm0.001$&$\pm0.001$&$\pm0.003$&$\pm0.002$&$\mathbf{\pm0.002}$&$\pm0.004$\\ \hline
$\Lambda$CDM&41&179&0&43&3&99&43&144& \textbf{1086}\\
&3\%&11\%&0\%&3\%&0\%&6\%&3\%&9\%&\textbf{66\%}\\
&$0.027$&$0.141$&$0.005$&$0.04$&$0.014$&$0.087$&$0.111$&$0.129$&$\mathbf{0.445}$\\
&$\pm0.002$&$\pm0.005$&$\pm0.0$&$\pm0.002$&$\pm0.001$&$\pm0.002$&$\pm0.002$&$\pm0.002$&$\mathbf{\pm0.006}$\\ \hline
\end{tabular}
\caption{The classification success matrix for the CNN.  The general structure of the table is the same as in table \ref{TAB::RESULTS::nn_cmatrix}. We see successful classifications of the two $f_{4}$ models, $f_{5}$, $f_{6}$ and \lcdm. However, large degeneracies remain within the neutrino mass variants of $f_{5}$ and $f_{6}$ gravity, respectively. In some cases, the wrong predictions can outnumber the correct ones as is the case for $f_{5}^{0.1}$ and $f_{6}^{0.06}$.}
\label{TAB::RESULTS::cnn_cmatrix}
\end{table*}
Compared to the fixed feature results in table \ref{TAB::RESULTS::nn_cmatrix} we find much larger true prediction values for many models. Exceptions are $f_{5}^{0.1}$ and $f_{6}^{0.06}$. Figure \ref{FIG::RESULTS::cnn_box} shows the statistics of the predictions for all classes in the test set and reveals that the $f_{6}^{0.06}$ and $f_{6}^{0.1}$ models, even as an ensemble, cannot be classified correctly by the CNN since the error bars on the medians of the predictions in their samples overlap with other $f_{6}$ models.
However, the degeneracy with \lcdm is now broken for all models and the CNN robustly discriminates most of the nine models from each other.

\subsection{Dependence on redshift}
\label{SEC::RESULTS:redshift}
A source redshift of \zsone is realistic for future space -and ground-based surveys but it is certainly optimistic for current ground-based surveys. On the other hand, it also does not test the full potential of our classification methods since one would expect a better classification accuracy for larger source redshifts. We therefore repeat training and classification for one lower ($z_{\mathrm{s}}=0.5$) and one higher ($z_{\mathrm{s}}=2$) source redshift. For simplicity we restrict this analysis to the CNN which delivered the best results. 

\begin{figure}
\includegraphics[width=.5\textwidth]{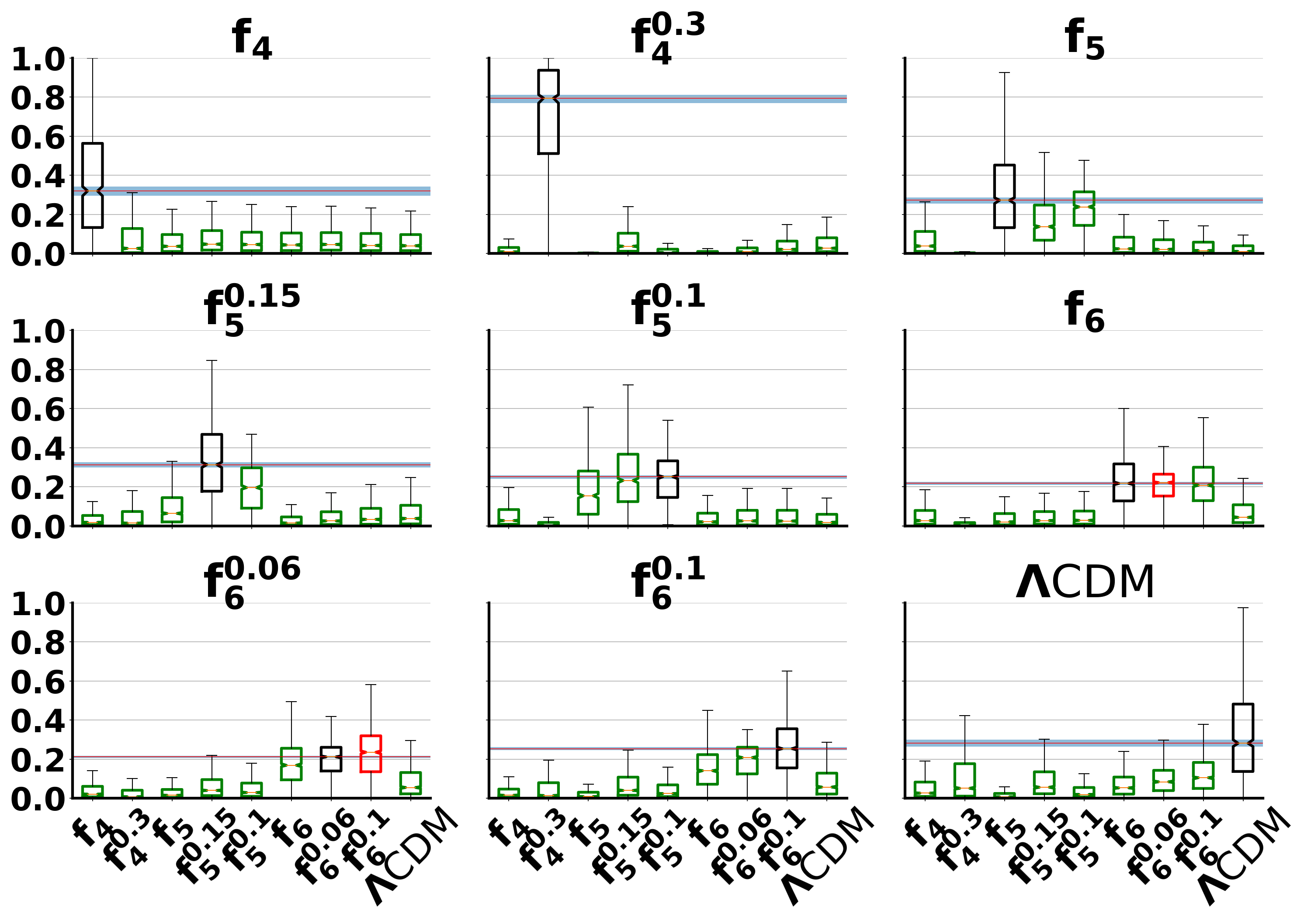}
\caption{Prediction statistics for the CNN and a source redshift of \zspfive. Compared to figure \ref{FIG::RESULTS::cnn_box}, the separation between the models becomes washed out. Two misclassifications occur: $f_{6}$ is incorrectly classified as $f_{6}^{0.06}$ and $f_{6}^{0.06}$ is misclassified as $f_{6}^{0.1}$. Also, the $f_{5}^{0.1}$ samples cannot be distinguished as an ensemble from the $f_{5}^{0.15}$ ones since their prediction medians overlap within the error bars.}
\label{FIG::RESULTS::cnn_z05_box}
\end{figure}

\begin{figure}
\includegraphics[width=.5\textwidth]{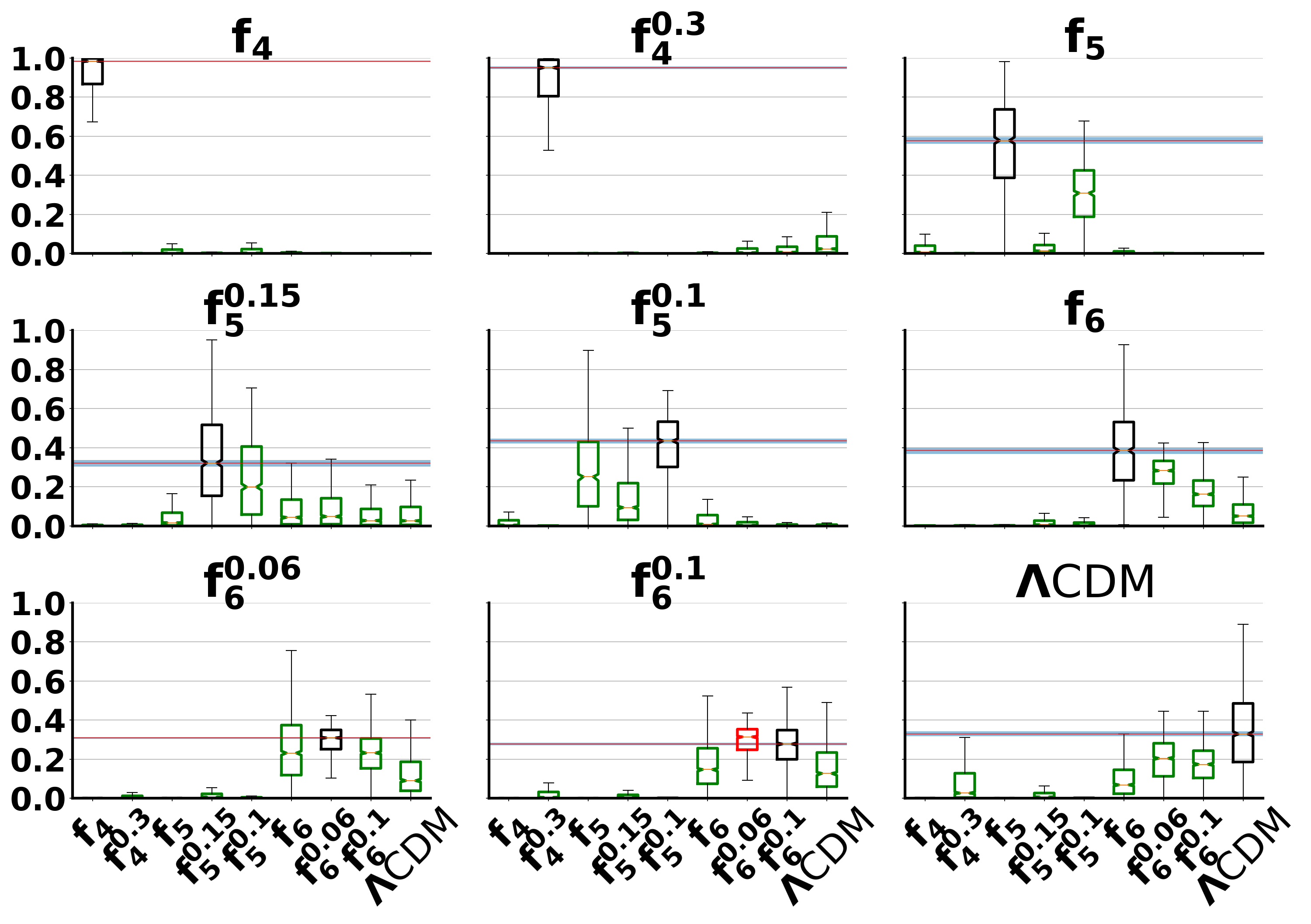}
\caption{Prediction statistics for the CNN and at a source redshift of \zstwo. Only the $f_{6}^{0.1}$ model remains degenerate given the error bar on the median of its sample predictions. In fact it is misclassified as $f_{6}^{0.06}$ by the CNN. }
\label{FIG::RESULTS::cnn_z20_box}
\end{figure}

For a source redshift \zspfive the overall accuracy drops significantly from 52\% to 44\%. When comparing the prediction statistics of the full set at this redshift in figure \ref{FIG::RESULTS::cnn_z05_box} with the reference at \zsone in figure \ref{FIG::RESULTS::cnn_box}, one can see that the decrease in the overall accuracy mainly stems from a weaker separation of the two $f_{4}$ models, $f_{5}$ and \lcdm. The known issue of degeneracies between the three neutrino masses for $f_{5}$ and $f_{6}$ are already present and more prominent. The issue of model misclassification for $f_{6}$ gravity gets worse with now two misclassifications.   
The improvements when going from \zsone to \zstwo are highlighted by figure \ref{FIG::RESULTS::cnn_z20_box}. 
For \zstwo the network's ability to distinguish between the base models increases and the overall classification accuracy is now 59\%. The  discrimination accuracy for massive neutrinos within each gravity model increases for the $f_{5}$ models and only the two $f_{6}$ models with massive neutrinos show significant overlap. Those models are also the only ones that show residual, but insignificant overlap with $\Lambda$CDM. Given the fact that the ensemble of $f_{6}^{0.1}$ maps also gets misidentified as $f_{6}^{0.06}$, it is clear that the discrimination within the $f_{6}$ models remains an issue even at a larger source redshift.

\begin{figure}
\includegraphics[width=.5\textwidth]{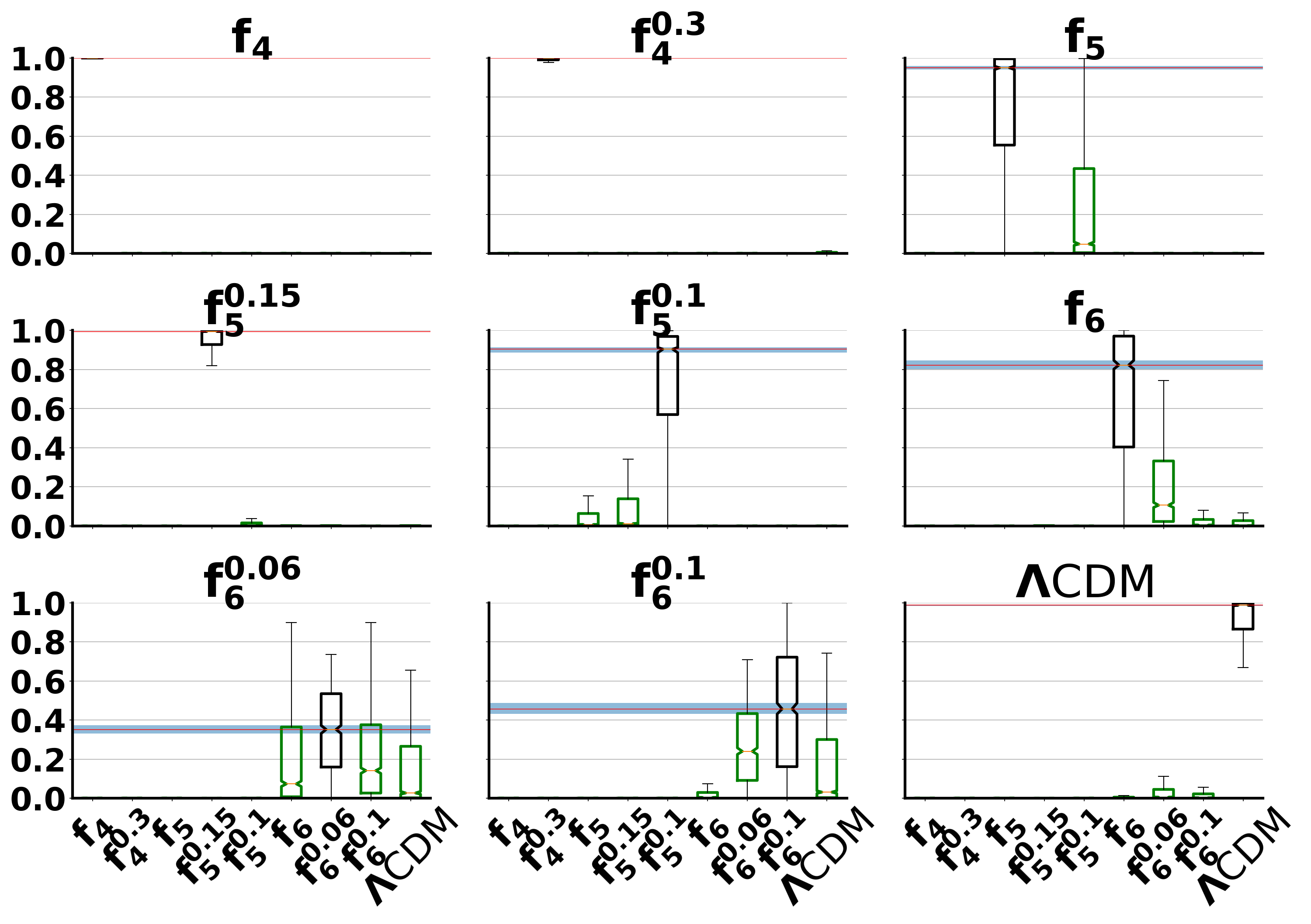}
\caption{Predictions statistics on the tomographic analysis with the CNN. For many classes the classification is so good, that the prediction samples cluster around the optimal value of 1. All models are correctly classified and within the error bars of the prediction medians, no model remains observationally degenerate. Only small similarities remain between the three $f_{6}$ models of varying neutrino mass.}
\label{FIG::RESULTS::cnn_tomo_box}
\end{figure}

As a last analysis using the CNN we perform a tomographic classification. For each line-of-sight realisation we are not using a single mass map at a specific source redshift but we feed data vectors of shape $\#x=(256,256,4)$ into the CNN where the four channels refer to $z_{\mathrm{s}}=0.5$, $z_{\mathrm{s}}=1$, $z_{\mathrm{s}}=1.5$ and $z_{\mathrm{s}}=2$, respectively. The classification success matrix for this analysis is shown in table \ref{TAB::RESULTS::cnn_tomo_cmatrix} and figure \ref{FIG::RESULTS::cnn_tomo_box} shows the familiar box-plot representation of the prediction-vector statistics. The overall classification success rate rises to 76\% and all models besides $f_{6}^{0.06}$ and $f_{6}^{0.1}$ now show correct classification rates of 74\% or clearly above. The probability of correctly classifying a single map in those two models are only 38\% or 50\%, respectively, however, a look at figure \ref{FIG::RESULTS::cnn_tomo_box} reveals that they are correctly classified as an ensemble and at high significance. Finally, it is worth noting that none of the models shows any degeneracy with \lcdm which is larger than 4\% according to table \ref{TAB::RESULTS::cnn_tomo_cmatrix}.    

\begin{table*}
\begin{tabular}{l|ccccccccc}
&$f_{4}$&$f_{4}^{0.3}$&$f_{5}$&$f_{5}^{0.15}$&$f_{5}^{0.1}$&$f_{6}$&$f_{6}^{0.06}$&$f_{6}^{0.1}$&$\Lambda$CDM\\
\hline
$f_{4}$& \textbf{1618}&0&7&0&10&3&0&0&0\\
&\textbf{99\%}&0\%&0\%&0\%&1\%&0\%&0\%&0\%&0\%\\
&$\mathbf{0.985}$&$0.0$&$0.007$&$0.0$&$0.006$&$0.002$&$0.0$&$0.0$&$0.0$\\
&$\mathbf{\pm0.002}$&$\pm0.0$&$\pm0.002$&$\pm0.0$&$\pm0.001$&$\pm0.001$&$\pm0.0$&$\pm0.0$&$\pm0.0$\\ \hline
$f_{4}^{0.3}$&0& \textbf{1501}&0&1&0&0&1&20&115\\
&0\%&\textbf{92\%}&0\%&0\%&0\%&0\%&0\%&1\%&7\%\\
&$0.0$&$\mathbf{0.91}$&$0.0$&$0.001$&$0.0$&$0.0$&$0.002$&$0.015$&$0.072$\\
&$\pm0.0$&$\mathbf{\pm0.006}$&$\pm0.0$&$\pm0.0$&$\pm0.0$&$\pm0.0$&$\pm0.001$&$\pm0.002$&$\pm0.005$\\ \hline
$f_{5}$&3&0& \textbf{1257}&2&375&1&0&0&0\\
&0\%&0\%&\textbf{77\%}&0\%&23\%&0\%&0\%&0\%&0\%\\
&$0.001$&$0.0$&$\mathbf{0.748}$&$0.002$&$0.247$&$0.001$&$0.0$&$0.0$&$0.0$\\
&$\pm0.001$&$\pm0.0$&$\mathbf{\pm0.008}$&$\pm0.001$&$\pm0.008$&$\pm0.001$&$\pm0.0$&$\pm0.0$&$\pm0.0$\\ \hline
$f_{5}^{0.15}$&0&0&1& \textbf{1470}&96&22&7&2&40\\
&0\%&0\%&0\%&\textbf{90\%}&6\%&1\%&0\%&0\%&2\%\\
&$0.0$&$0.0$&$0.001$&$\mathbf{0.873}$&$0.071$&$0.015$&$0.009$&$0.004$&$0.027$\\
&$\pm0.0$&$\pm0.0$&$\pm0.001$&$\mathbf{\pm0.007}$&$\pm0.005$&$\pm0.002$&$\pm0.001$&$\pm0.001$&$\pm0.003$\\ \hline
$f_{5}^{0.1}$&0&0&130&207& \textbf{1289}&12&0&0&0\\
&0\%&0\%&8\%&13\%&\textbf{79\%}&1\%&0\%&0\%&0\%\\
&$0.001$&$0.0$&$0.104$&$0.148$&$\mathbf{0.74}$&$0.008$&$0.0$&$0.0$&$0.0$\\
&$\pm0.0$&$\pm0.0$&$\pm0.005$&$\pm0.007$&$\mathbf{\pm0.008}$&$\pm0.002$&$\pm0.0$&$\pm0.0$&$\pm0.0$\\ \hline
$f_{6}$&0&0&0&15&1& \textbf{1206}&275&30&111\\
&0\%&0\%&0\%&1\%&0\%&\textbf{74\%}&17\%&2\%&7\%\\
&$0.0$&$0.0$&$0.0$&$0.01$&$0.001$&$\mathbf{0.676}$&$0.194$&$0.046$&$0.073$\\
&$\pm0.0$&$\pm0.0$&$\pm0.0$&$\pm0.002$&$\pm0.0$&$\mathbf{\pm0.008}$&$\pm0.005$&$\pm0.003$&$\pm0.005$\\ \hline
$f_{6}^{0.06}$&0&0&0&7&0&363& \textbf{627}&319&322\\
&0\%&0\%&0\%&0\%&0\%&22\%&\textbf{38\%}&19\%&20\%\\
&$0.0$&$0.0$&$0.0$&$0.005$&$0.0$&$0.226$&$\mathbf{0.344}$&$0.233$&$0.191$\\
&$\pm0.0$&$\pm0.0$&$\pm0.0$&$\pm0.001$&$\pm0.0$&$\pm0.007$&$\mathbf{\pm0.005}$&$\pm0.006$&$\pm0.007$\\ \hline
$f_{6}^{0.1}$&0&3&0&3&0&77&385& \textbf{827}&343\\
&0\%&0\%&0\%&0\%&0\%&5\%&24\%&\textbf{50\%}&21\%\\
&$0.0$&$0.002$&$0.0$&$0.002$&$0.0$&$0.064$&$0.272$&$\mathbf{0.455}$&$0.204$\\
&$\pm0.0$&$\pm0.001$&$\pm0.0$&$\pm0.001$&$\pm0.0$&$\pm0.004$&$\pm0.005$&$\mathbf{\pm0.008}$&$\pm0.008$\\ \hline
$\Lambda$CDM&0&1&0&6&0&57&69&37& \textbf{1468}\\
&0\%&0\%&0\%&0\%&0\%&3\%&4\%&2\%&\textbf{90\%}\\
&$0.0$&$0.001$&$0.0$&$0.006$&$0.0$&$0.039$&$0.059$&$0.044$&$\mathbf{0.85}$\\
&$\pm0.0$&$\pm0.001$&$\pm0.0$&$\pm0.001$&$\pm0.0$&$\pm0.003$&$\pm0.003$&$\pm0.003$&$\mathbf{\pm0.007}$\\ \hline
\end{tabular}
\caption{The classification success matrix for the tomographic analysis using the CNN. The general structure of the table is the same as in table \ref{TAB::RESULTS::nn_cmatrix}. We see good classification rates above 79\% and typically above 90\% for all models but the $f_{6}$-family. Also $f_{6}$ with vanishing neutrino mass is correctly classified 74\% of the time. The remaining degeneracies are limited to $f_{6}^{0.06}$ and $f_{6}^{0.1}$ with 38\% and 50\% classification accuracy, respectively, but given the error bars on the prediction mean the degeneracy is not significant for the ensemble of mass maps in the test set.}
\label{TAB::RESULTS::cnn_tomo_cmatrix}
\end{table*}

\subsection{Remarks on extracted features}
\label{SEC::RESULTS::features}
After presenting the raw classification results for different methods, we now briefly investigate what insight can be gathered into the actual meaning and importance of specific features that drive the classification success of different methods. To do so, we take a closer look at the training process. The first important observation is strikingly highlighted in table \ref{TAB::RESULTS::weights}, which shows that almost all of the most discriminating \wnd~features are Zernike coefficients derived from the Fourier transform of the raw image or from the Fourier transform of the edge -or wavelet-processed image. This is interesting since Zernike polynomials were originally introduced to describe the effects of certain optical elements such as lenses or reflecting surfaces in optical imaging \citep{Zernike1934}. This suggests that a decomposition of mass maps into a function set that has a well-defined physical meaning does indeed lead to a good general representation of our data. In addition, all those features are derived from transformations of the raw mass map which shows the power of filtering the input data as e.g. shown by \citet{Peel2018}. The ranking of the standard features shown in table \ref{TAB::RESULTS::classicweights} is less dominated by a single class, although the power spectrum and peak counts seem most relevant. The good results with a neural network as classifier shows that the optimal combination of such classical features leads to a good classification even without the need for additional descriptors.

\begin{figure}
\begin{center}
\includegraphics[width=.45\textwidth]{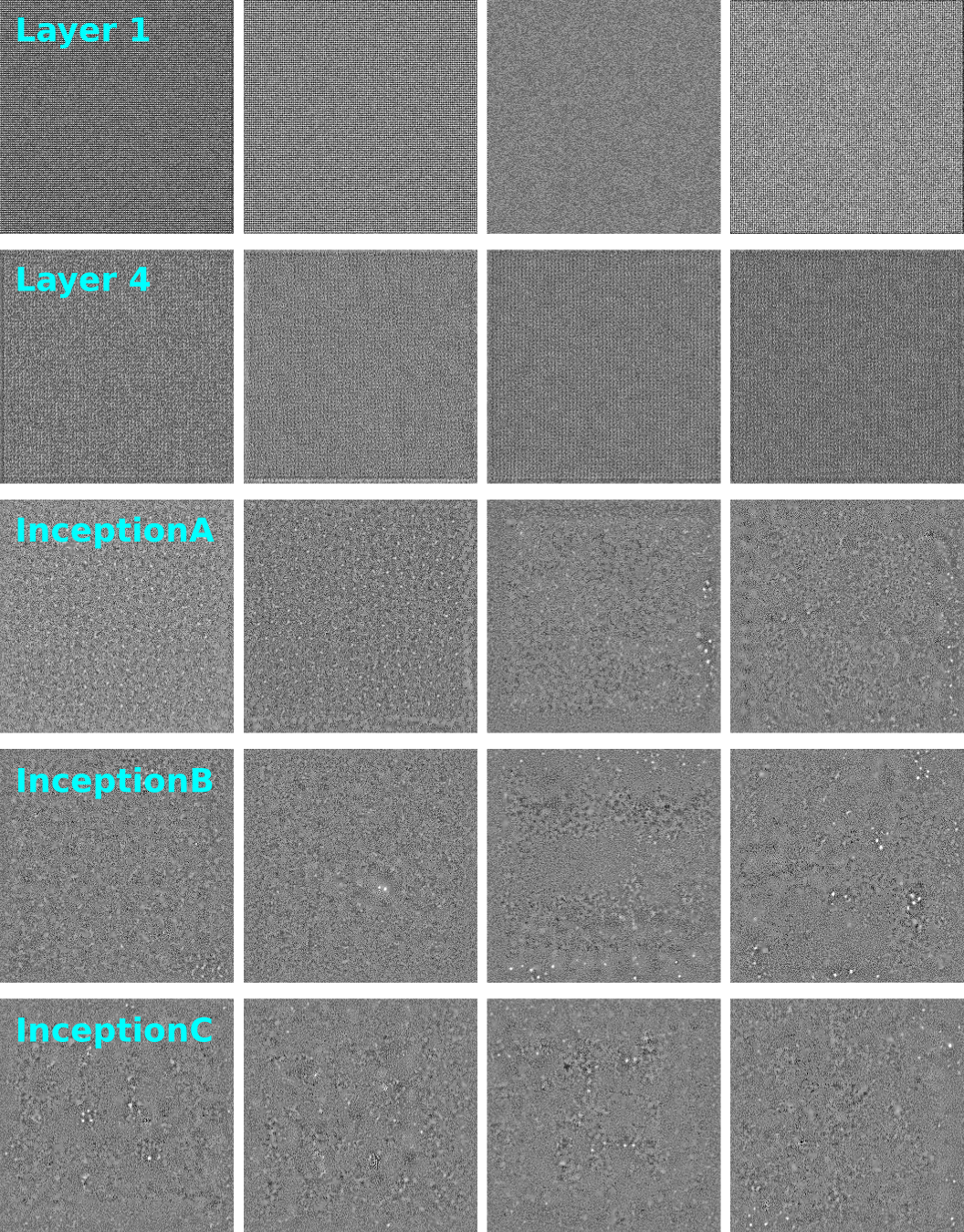}
\end{center}
\caption{Visualisations of the convolutional filters applied by the CNN at different depths of the network.}
\label{FIG::RESULTS::filter_depth}
\end{figure}

CNNs often deliver superior results compared to other methods for certain tasks, but it is often believed that they are harder to understand and interpret. We are attempting to dissolve this believe by applying visualisation techniques for the different filters linked together in a deep neural network \citep{Girshick2013,Zeiler2013,Szegedy2013,Springenberg2014} and in order to reveal the inner workings of the complex model. We follow the approach of \citet{Simonyan2013} to extract our filter responses\footnote{\flushleft{Also see \href{https://github.com/keras-team/keras/blob/master/examples/conv_filter_visualization.py}{https://github.com/keras-team/keras/blob/master/examples/conv\_filter\_visualization.py}}}. Starting from an image of random numbers with the same shape as our mass maps, we retrieve the output of every convolutional layer in the network and perform a gradient ascent in order maximise the response of those layers. While this is of course not a unique solution, the result of the final iteration of the ascent represents an example which triggered a strong response at a particular depth in the network. In figure \ref{FIG::RESULTS::filter_depth} we show a few examples. The top row shows the four channels which had the strongest loss compared to the initial random image in CNN layer one. That is the $3\times 3$ convolution marked with index 2 in table \ref{TAB::RESULTS::cnn}. The second row shows the top four channel responses of the $3\times 3$ convolution with stride two just above the input layer in figure \ref{FIG::RESULTS::stem}. The row marked with InceptionA shows the most responsive channels among all four convolutions just below the concatenation layer in figure \ref{FIG::RESULTS::stem} and equivalently for the figure rows marked InceptionB and C.    
As is typical for CNNs \citep{Zeiler2013}, the very first level extracts very regular horizontal and vertical stripe patterns from the image. The stripes turn into a grid pattern deeper into the network and once arriving at the end of the InceptionA layer we can identify patterns of peaks and troughs which are either grouped regularly or along larger structures. It is not surprising that the earlier layers of the network, up to InceptionA, perform a global filtering of the map that highlights structure as long as the image still consists of a relatively large number of pixels. It is just from the finer InceptionB layers on that more specific structures, like objects that look like individual clusters or voids are picked up. It is such detailed analyses of the inner structure of trained CNNs that will lead to a deeper understanding why those networks work so well. This can potentially lead to the development of more specific algorithms at lower numerical cost but with similar or better classification performance. 

\section{Conclusions}
\label{SEC::CONCL}
We studied the ability of different kinds of machine learning techniques to discriminate between highly degenerate cosmological models, which combine the effects of modified gravity and massive neutrinos on structure formation. For this purpose we used a subset of the \dustp simulation suite which consists of \lcdm and eight $f(R)$ models of gravity in the range of $-1\times 10^{-4}
\le f_{R0} \le -1\times 10^{-6}$. The neutrino masses in the simulations span $0~{\rm eV} \le m_{\nu  } \le 0.3~{\rm eV}$.  Lensing convergence maps produced from these simulations provided the input for the different classification methods.

In order to characterise the mass maps we used three different approaches to feature extraction. Commonly used statistics in astrophysics such as, and among others, the power spectrum, peak counts and Minkowski functionals were combined into a single feature vector. In order to probe features which are more common to the field of computer vision and digital image processing, we used the publicly available \wnd~algorithm which produces a large feature vector that combines a variety of common and more exotic descriptors and statistics. As the most flexible method of feature extraction we used a convolutional neural network (CNN). For classification we tested a nearest-neighbour method in feature space and a fully-connected neural network.

We provide an overview of the classification results from section \ref{SEC::RESULTS} in table \ref{TAB::CONCL::summary} and our results can be summarised as follows:
\begin{table*}
\begin{tabular}{ccccccc}
$\Theta$&$\zeta$&$z_{\mathrm{s}}$&total accuracy&degenerate classes&\lcdm performance&Reference\\
\hline
classic&nearest neighbour&$1.0$&22\%&8&14\%/15\% ($f_{4}^{0.3}$)&repository\\
\wnd&nearest neighbour&$1.0$&25\%&7&24\%/24\%($f_{4}^{0.3}$)&repository\\
classic&neural network&$1.0$&39\%&3&47\%/16\%($f_{4}^{0.3}$)&Tab.~\ref{TAB::RESULTS::nn_cmatrix},~Fig.~\ref{FIG::RESULTS::nn_box}\\
\wnd&neural network&$1.0$&36\%&4&42\%/24\%($f_{4}^{0.3}$)&repository\\
CNN&neural network&$0.5$&44\%&3&52\%/15\%($f_{4}^{0.3}$)&Fig.~\ref{FIG::RESULTS::cnn_z05_box}\\
CNN&neural network&$1.0$&52\%&2&66\%/11\%($f_{4}^{0.3}$)&Tab.~\ref{TAB::RESULTS::cnn_cmatrix},~Fig.~\ref{FIG::RESULTS::cnn_box}\\
CNN&neural network&$2.0$&59\%&1&53\%/12\%($f_{4}^{0.3}$)&Fig.~\ref{FIG::RESULTS::cnn_z20_box}\\
CNN&neural network&$0.5,1.0,1.5,2.0$&76\%&0&90\%/4\%($f_{6}^{0.06}$)&Tab.~\ref{TAB::RESULTS::cnn_tomo_cmatrix},~Fig.~\ref{FIG::RESULTS::cnn_tomo_box}\\
\end{tabular}
\caption{A summary of the performance of the different methods used in this analysis. $\Theta$ indicates the feature extraction function as described in section \ref{SEC::METHOD::features}. $\zeta$ is the classification function introduced in section \ref{SEC::METHOD::class} and $z_{\mathrm{s}}$ is the convergence map source redshift. Degenerate classes is the number of all models for which the median and its error for the predictions of the true test set class overlaps with the median and its error of the predictions for any other class. The table also lists the performance for the particularly important \lcdm class and shows the classification accuracy of each method for this model as well as the largest misclassification rate and the associated model. A reference to the detailed results of each models is given in the last column, where the reference 'repository' points to the online repository mentioned in section \ref{SEC::METHOD::setup}.}
\label{TAB::CONCL::summary}
\end{table*}
\begin{enumerate}
\item Nearest-neighbour classifiers based on distances in feature space are not delivering robust results. No matter if a small classical feature vector is used or a longer version based on computer vision, the total classification accuracy stays below 25\%. Eight, out of the nine tested models, remain observationally degenerate\footnote{We declare a model as degenerate if the median and its error for the predictions of a true test set class overlaps with the median and its error of the predictions for any other class.}.
\item With the same classical or computer vision feature vectors, a neural network delivers a much more robust classification than the nearest-neighbour method. The total success rate for the classical feature vector is 39\% and the number of degenerate models reduces to three.
\item The longer feature vector containing 2919 features inspired by computer vision delivers a slightly worse classification of our models than the shorter vector with 99 classical descriptors. The total classification success rate is 3\% lower and the method produces one additional degenerate model. Some of the computer vision feature may very well be useful, but currently we see no advantage of using features inspired by digital image processing compared to features well-established in cosmology.    
\item A CNN delivers the best classification results with 52\% correct classifications at source redshift $z_{\mathrm{s}}=1.0$. The number of degenerate models reduces to two, both of which are part of the same $f_{6}$ model of gravity.  
\item Classification success is clearly a function of mass map source redshift. While going from \zsone to \zspfive the success rate of the CNN decreases by 8\% and the number of degenerate models increases by one. When going from \zsone to \zstwo the accuracy increases by 7\% and the number of degenerate models reduces by one. This increase of success rate with increasing redshift is not surprising since more information relevant to structure formation can be picked up along a deeper line-of-sight.  
\item When using a CNN in a tomographic analysis of four different mass map source redshifts along the same line-of-sight all observational degeneracies are fully broken. The total classification success rate increases to 76\%. 
\end{enumerate}

A number of improvements to our methodology come to mind and we reserve them for future work. Firstly, the flexible features derived by a CNN can be combined with fixed features that are known to contribute to a successful classification of degenerate models. Secondly, instead of working on the raw image data, a clever transformation can be applied to the input data to enhance features that allow for the desired discrimination. We attempt such an approach in the context of machine learning in Peel et al. (2018b, PRL submitted). In fact, the CNN used in this work applies such transformations as we discussed in section \ref{SEC::RESULTS::features}. A careful analysis of the filtering process of a CNN at the early levels of its filter chain can provide useful insight into the most powerful image transformation for a given classification task. Furthermore, the careful analysis of the filters at a much deeper level of the network might actually lead to more insights on structure formation in different models, since it is at this deeper level where individual structure is characterised and isolated by the algorithm. 

Much work is left to be done before this machine learning approach to the classification of mass maps in different cosmological models can be applied to real data. In this work we limited ourselves to optimal noise-free maps in order to see how different methodologies compare under optimal conditions. The influence of pixel shot-noise, observational systematics and practical issues like masking and image artefacts needs to be studied in detail. Furthermore, since the currently most successful methods use a supervised training process with labelled data based on numerical simulations, it needs to be carefully investigated how closely those simulated maps resemble a real observation. Without this important sanity check, even the best machine learning technique is useless since it learns the wrong data. 

\section*{Acknowledgements}
We would like to thank Ofer Springer for useful discussions about deep learning.  JM has received funding from the European Union's Horizon 2020 research and Innovation programme under the Marie Sk\l{}odowska-Curie grant agreement No 664931. AP acknowledges support from an Enhanced Eurotalents Fellowship, a Marie Sk\l{}odowska-Curie Actions Programme co-funded by the European Commission and Commissariat {\`a} l'{\'e}nergie atomique et aux {\'e}nergies alternatives (CEA). CG and MB acknowledge support from the Italian Ministry for Education, University  and  Research  (MIUR)  through the  SIR  individual  grant SIMCODE (project number RBSI14P4IH).  CG and MM acknowledge support from the Italian  Ministry of  Foreign Affairs  and International  Cooperation, Directorate General  for Country  Promotion (Project "Crack the lens").   We also  acknowledge the support  from the  grant  MIUR PRIN  2015  "Cosmology and  Fundamental Physics:  illuminating  the  Dark   Universe  with  Euclid";  and  the financial contribution from the agreement ASI n.I/023/12/0 "Attivit\`a relative  alla  fase  B2/C  per   la  missione  Euclid". The \dustp simulations analyzed in this work have been performed on the Marconi supercomputing machine at Cineca thanks to the PRACE project SIMCODE1 (grant  nr.  2016153604)  and  on  the  computing  facilities  of the  Computational  Center  for  Particle  and  Astrophysics (C2PAP) and of the Leibniz Supercomputer Center (LRZ) under the project ID pr94ji.We gratefully acknowledge the support of NVIDIA Corporation with the donation of the two Titan Xp GPUs used for this research.







\appendix
\section{Wnd-charm features}
\label{SEC::APP1}
The total length of the \wnd~feature vector entails 2919 descriptors, which can be divided into five families. We provide an overview of the features and their respective families in table \ref{TAB::APP1::features}. The algorithm does not only work on the image itself (raw), but also on its Fourier (F), Wavelet (W), Chebyshev (C) or Edge transformation (E) as indicated by the 'Input' column of table \ref{TAB::APP1::features}. Transformations of transformations are considered by the bracket notation. While Fourier and Chebyshev transforms are implemented using common algorithms and methodologies, the Wavelet transformation is performed with a one level filter pass with a 5th order symlet \citep{Orlov2008} and the Edge transformation is carried out using a Prewitt operator \citep{Prewitt1970a} to approximate the image gradient. 

The pixel statistics family is made out of four different subclasses, with the simplest being the intensity statistics consisting of mean, median, standard deviation, minimum and maximum. The multi-scale histograms are calculated by using three, five, seven or nine bins to order the pixel amplitudes. The counts in each of those bins makes up the 24 features in this subclass. The combined moments are mean, standard deviation, skewness and kurtosis, which are calculated in a horizontal stripe through the image centre and with a width which is half the total image width. The stripe is then rotated by 45, 90 and 135 degrees and the measurement is repeated. Those 16 numbers are sampled into three bins each, providing a total of 48 features. The Gini coefficient \citep{Abraham2003} is a measure of how equal the spectrum of pixel intensities is distributed within the image.

The second feature family is comprised of polynomial decompositions. The coefficients of an order 20 Chebyshev  and an order 23 Chebyshev-Fourier \citep{Orlov2006} transformation are sorted into 32 bin histograms. Radon transformations are carried out along lines with an inclination angle of 0, 45, 90 and 135 degrees with respect to the image horizontal \citep{Radon1917} and ordered in 3 bin histograms. The class of Zernike coefficients is derived from a 2D Zernike decomposition of the image \citep{Teague1980} and the first 72 of those coefficients contribute to the feature vector. 

The use of textures is common in image processing and is a way of describing spatial correlations of intensity values. We extract seven Gabor filters \citep[e.g.][]{Fogel1989} using Gaussian harmonic functions and define their image occupation area as a feature. Tamura textures are described in detail in \citet{Tamura1978} and \wnd~uses contrast, directionality, coarseness sum and coarseness binned into a 3 sample histogram. The 28 Haralick textures are specific properties of the grey-level dependence matrix of the image and are described in \citet{Haralick1973}. The fractal analysis is based on a Brownian motion model of the image following \citet{Wu1992} and \wnd~uses the first 20 parameters of this analysis as features. 

Object statistics are only derived from the raw image data. The starting point is an edge transform using a Prewitt filter and mean, median, variance and 8-bin histogram of both image gradient and its directionality add up to 22 features, which are supplemented by the total number of edge pixels, their genus and the differences between the directionality bins. Otsu features and their inverse are calculated after the application of an Otsu threshold \citep{Otsu1979}. Finally, for all objects the algorithm calculates minimum, maximum, mean, median, variance and 10-bin histogram for area and image-centre distance of all Otsu objects in the image.  

\begin{table*}
\begin{tabular}{cccccc}
\hline
Family & Class & Features & Input & Reference\\
\hline
\hline
Pixel& Combined moments & 48 & raw, F, W, C, C(F), W(F) & -- \\
statistics&&&F(W), F(C), C(W), E, F(E), W(E) & \\
& Gini coefficient & 1 & raw, F, W, C, C(F), W(F) & \citet{Abraham2003} \\
&&&F(W), F(C), C(W), E, F(E), W(E) & \\
& Multiscale histograms & 24 & raw, F, W, C, C(F), W(F) & -- \\
&&&F(W), F(C), C(W), E, F(E), W(E) & \\
& Pixel intensity statistics & 5 & raw, F, W, C, C(F), W(F) & -- \\
&&&F(W), F(C), C(W), E, F(E), W(E) & \\
\hline
Polynomial& Chebyshev coefficients & 32 & raw, F, W, C, F(W), E,
F(E), W(E) &--\\
decomposition& Chebyshev-Fourier coefficients & 32 & raw, F, W, C, F(W), E, 
F(E), W(E) &
\citet{Orlov2006}\\
& Radon coefficients & 12 & raw, F, W, C, C(F), W(F) & \citet{Radon1917}\\
&&&F(W), F(C), C(W), E, F(E), W(E) & \\
& Zernike coefficients & 72 & raw, F, W, C, F(W), E, F(E), W(E)&
\citet{Teague1980}) \\
\hline
Textures& Fractal analysis & 20 & raw, F, W, C, C(F), W(F) & \citet{Wu1992} \\
&&&F(W), F(C), C(W), E, F(E), W(E) & \\
& Gabor & 7 & raw & \citet{Fogel1989} \\
& Haralick & 28 & raw, F, W, C, C(F), W(F)&\citet{Haralick1973}\\
&&&F(W), F(C), C(W), E, F(E), W(E) & \\
& Tamura & 6 &	 raw, F, W, C, C(F), W(F)& \citet{Tamura1978} \\
&&&F(W), F(C), C(W), E, F(E), W(E) & \\
\hline
Objects& Edge features & 28 & raw & \citet{Prewitt1970a} \\
& Otsu object features & 34 & raw & \citet{Otsu1979}\\
& Inverse Otsu object features & 34 & raw & \citet{Otsu1979}\\
\end{tabular}
\caption{\wnd~image features used in this analysis.}
\label{TAB::APP1::features}
\end{table*}

\section{Deep neural networks}
\label{SEC::APP2}
In this appendix we collect some more detailed information about deep neural networks. The first section formally defines all network layers used in this work and the second section deals with activation functions. The third section  provides a thorough description about the architecture of the convolutional neural network (CNN) that we use in our analyses. 
\subsection{Layers}
\label{SEC::APP2::layers}
Given a 2D input vector $I_{ijc}$ with $\#I_{ijc}=(X,Y,l)$ a convolution layer applies the following operation to produce an output $O_{ijc}$ 
\begin{align}
\label{EQU::METHOD::conv}
&\mathrm{Conv}(n,m,\Delta i,\Delta j,p,C)I_{i^{\prime}j^{\prime}c^{\prime}} = O_{ijc}  \\
&O_{ijc} = B_{c}+\sum\limits_{i^{\prime}=1}^{n}\sum\limits_{j^{\prime}=1}^{m}\sum\limits_{c^{\prime}=1}^{l} W_{i^{\prime} j^{\prime} c^{\prime}}^{c} I_{(i\Delta i+i^{\prime})(j\Delta j+j^{\prime}) c^{\prime}} \\
&w_{\mathrm{conv}} = \left\{ B_{c}, W_{ijc}\right\} \quad \forall c \\
&\#O_{ijc} = \left(\frac{X}{\Delta i},\frac{Y}{\Delta j},l\right) \quad \mathrm{for} \quad p =\mathrm{s} \\
&\#O_{xyc} = \left(\frac{X}{\Delta i}-\frac{n}{2},\frac{Y}{\Delta j}-\frac{m}{2},C\right) \quad \mathrm{for} \quad p =\mathrm{v}.
\end{align}
The stride parameters $\Delta i$ and $\Delta j$ allow one to implement dimensional reduction. The parameter $p$ indicates if the input data is padded, which means that additional rows and columns are added in order to produce an output which has exactly the same spatial shape as the input, at least in the absence of stride. This is known as \textit{same} padding $p=\mathrm{s}$. Alternatively the data can stay unaltered, or \textit{valid} $p=\mathrm{v}$, which means that the spatial dimensions of the data vector are slightly reduced, since every convolution must be fully contained within the 2D image domain.

We use four different kinds of pooling layers. Their main functionality is either an averaging 
\begin{align}
\label{EQU::METHOD::avgppol}
&\mathrm{AvgPool}(n,m,\Delta i,\Delta j,p)I_{i^{\prime}j^{\prime}c} = O_{ijc}  \\
&O_{ijc} = \frac{1}{nm}\sum\limits_{i^{\prime}=1}^{n}\sum\limits_{j^{\prime}=1}^{m}I_{(i\Delta i+i^{\prime})(j\Delta j+j^{\prime})c} \\
&\#O_{ijc} = \left(\frac{X}{\Delta i},\frac{Y}{\Delta j},C\right) \quad \mathrm{for} \quad p =\mathrm{s} \\
&\#O_{ijc} = \left(\frac{X}{\Delta i}-\frac{n}{2},\frac{Y}{\Delta j}-\frac{m}{2},l\right) \quad \mathrm{for} \quad p =\mathrm{v},
\end{align}
or a maximum selection operation
\begin{align}
\label{EQU::METHOD::maxppol}
&\mathrm{MaxPool}(n,m)I_{i^{\prime}j^{\prime}c} = O_{ijc}  \\
&O_{ijc} =\max\left\{I_{(i\Delta i+i^{\prime})(j\Delta j+j^{\prime})c}\right\}_{i^{\prime}=1,j^{\prime}=1}^{n,m}\\
&\#O_{ijc} = \left(\frac{X}{\Delta i},\frac{Y}{\Delta j},l\right) \quad \mathrm{for} \quad p =\mathrm{s} \\
&\#O_{ijc} = \left(\frac{X}{\Delta i}-\frac{n}{2},\frac{Y}{\Delta j}-\frac{m}{2},l\right) \quad \mathrm{for} \quad p =\mathrm{v}.
\end{align}
Both pooling layers exist also as global versions, indicated by GlobalMaxPool and GlobalAvgPool, where all entries in a channel are considered for either the maximum or averaging operation. In this case the shape of the output reduces to $(1,1,l)$.

A concatenation layer performs a stacking operation along the $c$-axis, which means that the spatial dimensionality of each input $I_{ijc}$ must be the same $(X,Y)$.  
\begin{align}
\label{EQU::METHOD::concat}
&\mathrm{Concatenate}(I_{ijc_{1}},...,I_{ijc_{n}}) = O_{ijc}  \\
&O_{ijc} = I_{ijc_{1}} \oplus ... \oplus I_{ijc_{n}} \\
&\#O_{ijc} = \left(X,Y,C_{1}+...+C_{n}\right),
\end{align}
where the $\oplus$ operator implements the channel stacking. The respective number of input channels is $C_{1},...,C_{n}$ for a concatenation of $n$ layers.

Fully-connected, sometimes called affine, layers create a linear mapping between the input and the output
\begin{align}
\label{EQU::METHOD::fc}
&\mathrm{FC}(n)I_{ijc^{\prime}} = O_{ijc} \\
&O_{ijc^{\prime}} = B_{ij}+\sum\limits_{k=1}^{X}A_{c^{\prime}k}I_{ijk} \\
&w_{\mathrm{FC}} = \left\{B_{ij},A_{c^{\prime}k} \right\} \\
&\# O_{ij} = (1,1,n).  
\end{align}
Here, we assume that the input layer has a simple 1D shape (1,1,X).

\subsection{Activation functions}
\label{SEC::APP2::acti}
We use three kinds of activation functions. Feature extraction layers such as convolution and pooling layers are often followed by rectangular linear units (ReLU) or its generalisation which is commonly called a leaky rectangular linear unit
\begin{align}
\label{EQU::METHOD::activation}
&\mathrm{ReLU}(x) = \max(0,x) \\
&\mathrm{leakyReLU}(x;\alpha) = 
\begin{cases}
x \quad &x\geq 0 \\
\alpha x \quad &\mathrm{otherwise.}
\end{cases}
\end{align}
The last fully connected layer in a neural network that is used for classification is often followed by softmax function in order to produce predictions in the final output of the network 
\begin{equation}
\label{EQU::METHOD::softmax}
\text{Softmax}(x)_{n} = \frac{\exp{x}_{n}}{\sum\limits_{j=1}^{N}\exp{x_{j}}} \qquad \text{for}\quad n=1,...,N.
\end{equation}

\subsection{CNN architecture}
\label{SEC::APP2::arch}
\begin{figure}
\centering
\includegraphics[height=.4\textheight]{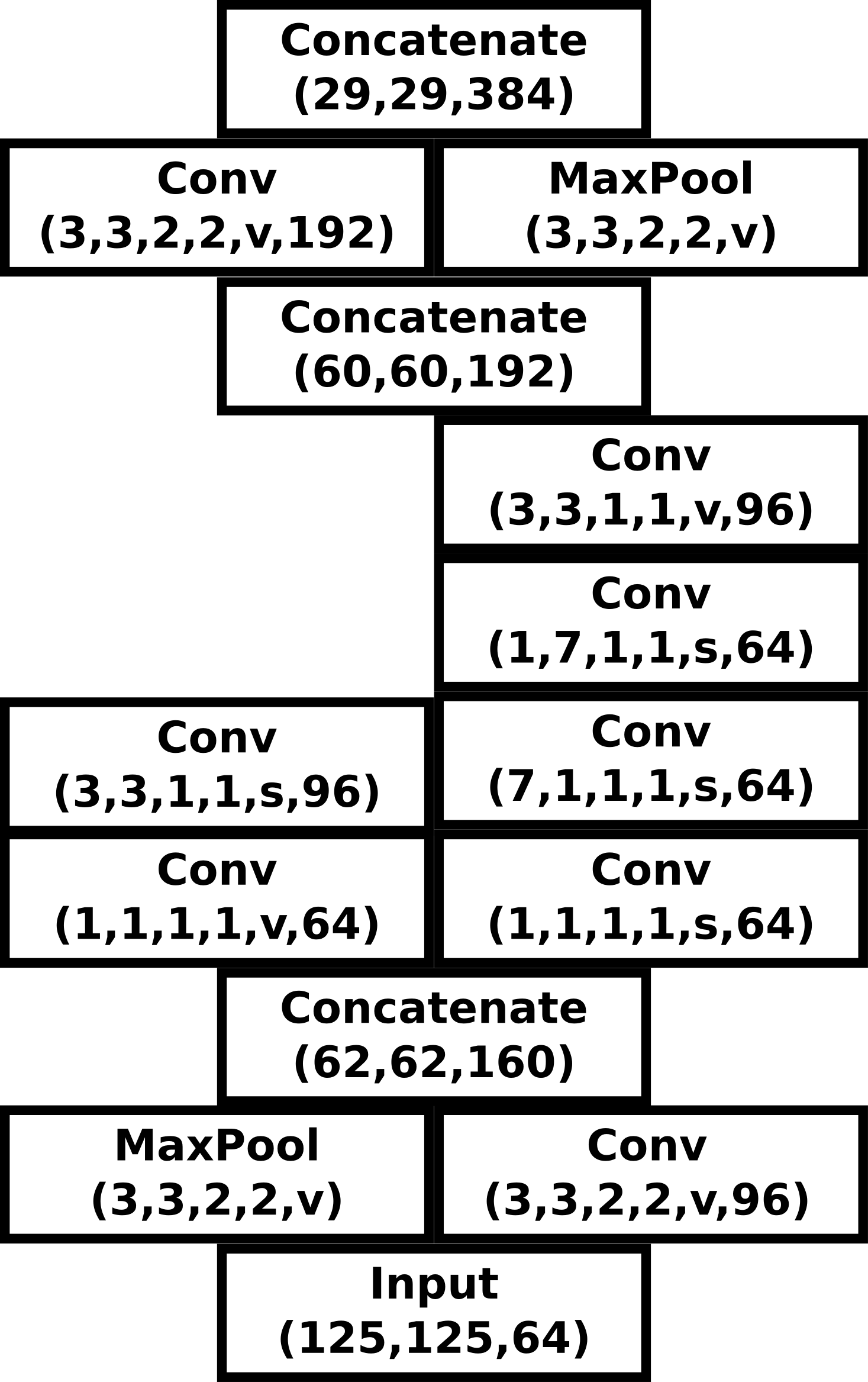}
\caption{The internal structure of the stem Inception layer. The layout is identical to \citet{Szegedy2016}, but with the image dimensions of our mass maps.}
\label{FIG::RESULTS::stem}
\end{figure}

In section \ref{SEC::RESULTS::cnn} we described the global structure of our CNN, which is largely based on \citet{Szegedy2016}. Here we describe in detail the purpose of each of the functional elements that are shown in table \ref{TAB::RESULTS::cnn}. After three conventional $3\times3$ convolutions for initial feature extraction and dimensional reduction, we enter the StemInception layer, which is shown in detail in figure \ref{FIG::RESULTS::stem}. In our CNN, the purpose of the stem layer is twofold. Firstly, it further reduces the data vector from $125\times125$ pixels down to $29\times29$ pixels, which is a computationally manageable size for applying a large number of convolution channels. Secondly, it already applies a more refined combination of $3\times3$, $7\times7$ and  $1\times1$ convolutions. The latter only have the purpose of channel reduction as explained in \citet{Szegedy2016}. 
The stem layer is followed by the three main inception layers A, B and C. The main purpose of those layers is feature extraction, with a particularly large number of convolutions of varying kernel size. Between the main feature extraction layers we insert reduction layers, breaking the image up further into smaller postage stamps and allowing the application of a larger number of convolution channels within acceptable runtimes and within the memory constraints of the hardware we deploy.
The very last concatenation layer of InceptionC is followed by a global averaging layer and a single fully connected layer for classification. 

\begin{figure}
\includegraphics[width=.45\textwidth]{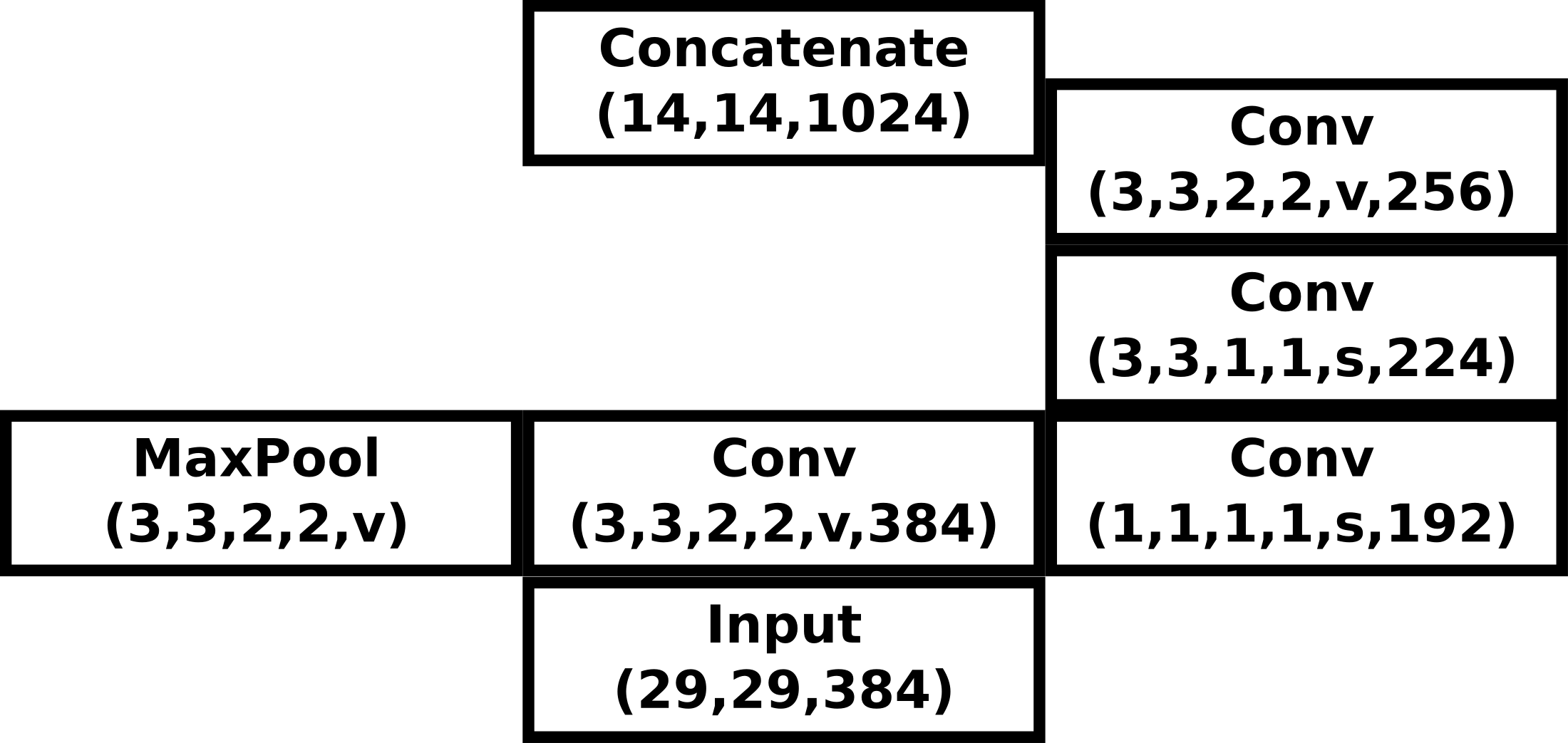}
\caption{ReductionA layer of our CNN, based on \citet{Szegedy2016}.}
\label{FIG::RESULTS::ra}
\end{figure}

\begin{figure}
\includegraphics[width=.45\textwidth]{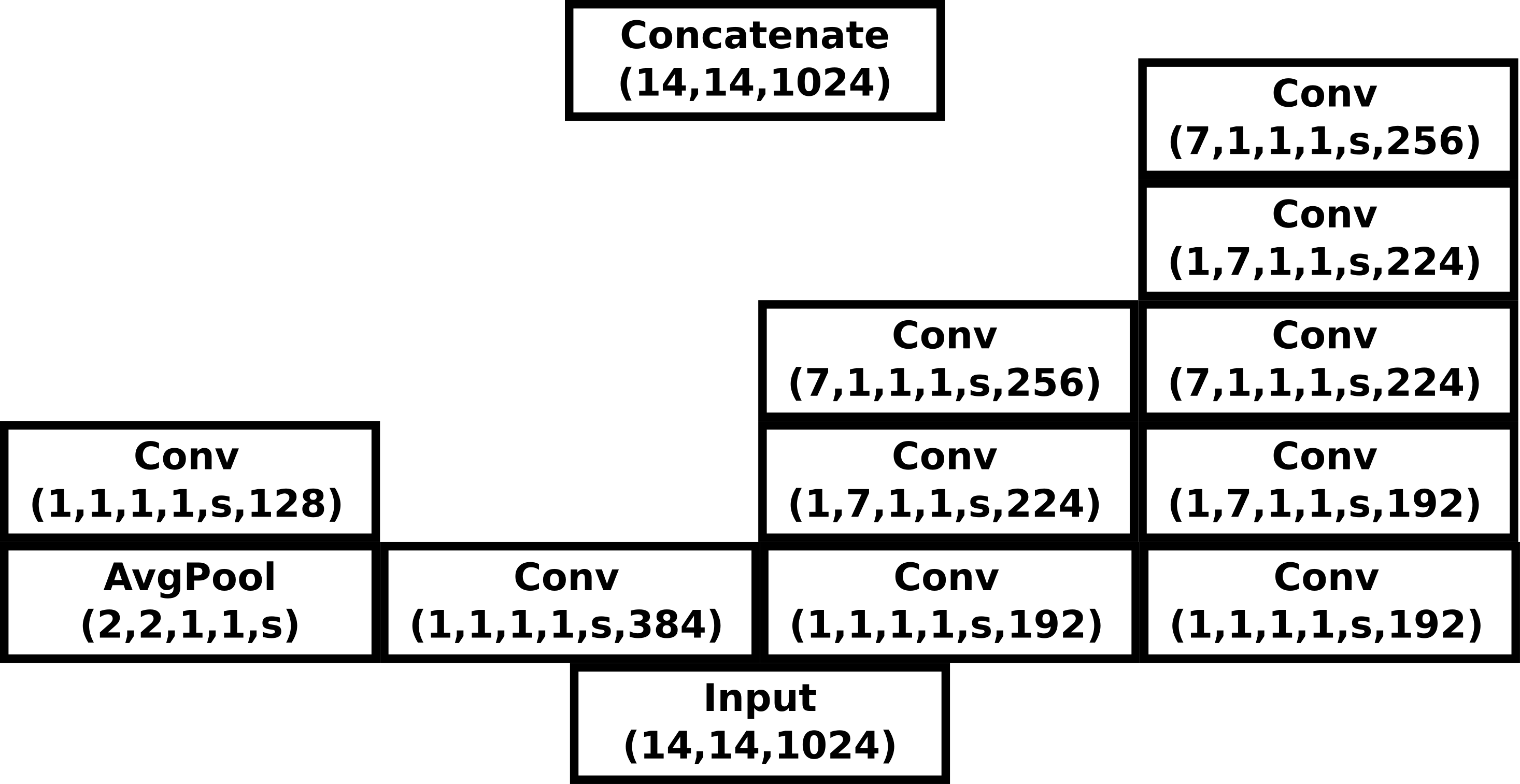}
\caption{InceptionB layer of our CNN, based on \citet{Szegedy2016}.}
\label{FIG::RESULTS::b}
\end{figure}

\begin{figure}
\includegraphics[width=.45\textwidth]{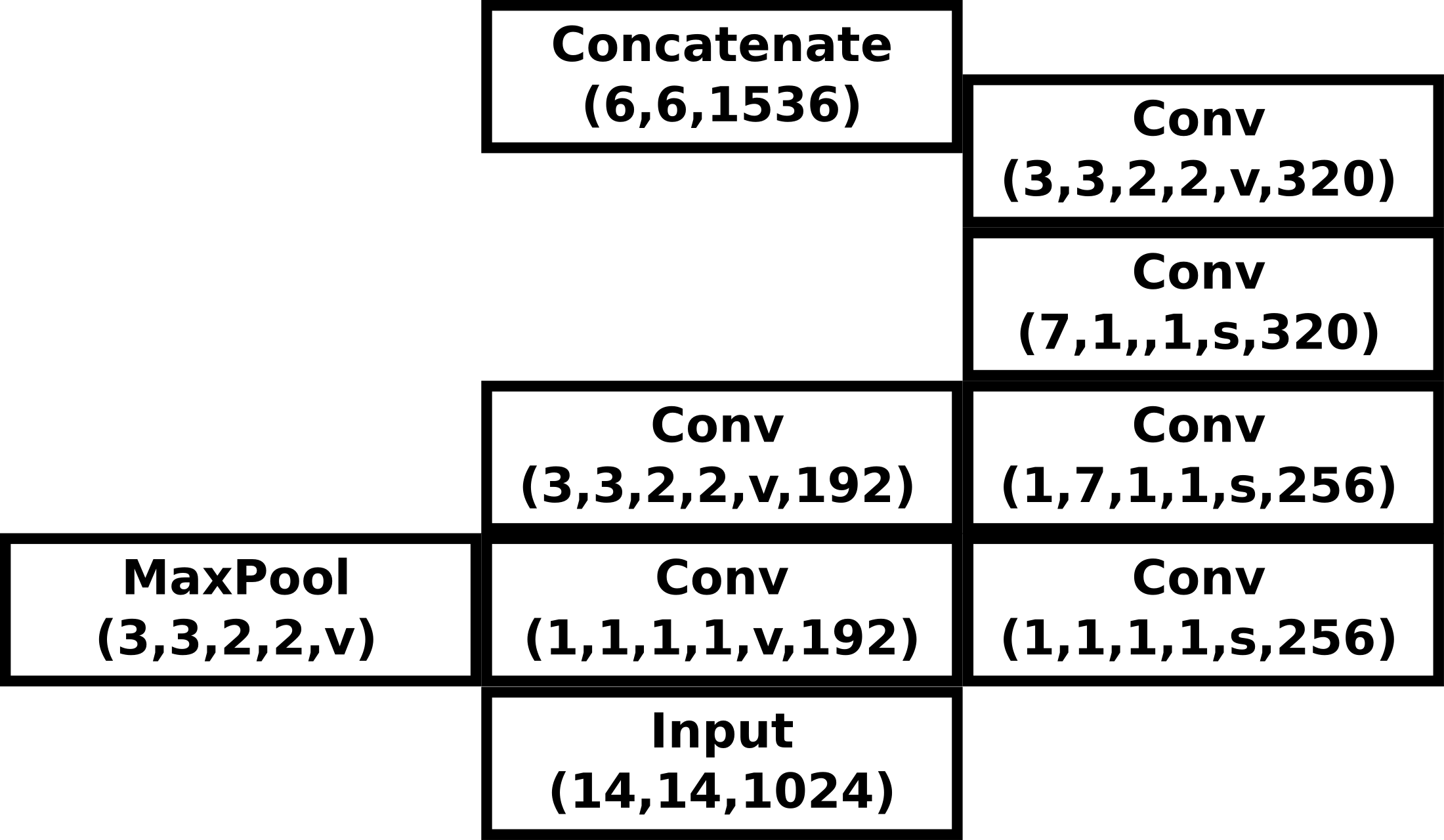}
\caption{ReductionB layer of our CNN, based on \citet{Szegedy2016}.}
\label{FIG::RESULTS::rb}
\end{figure}

\begin{figure}
\includegraphics[width=.45\textwidth]{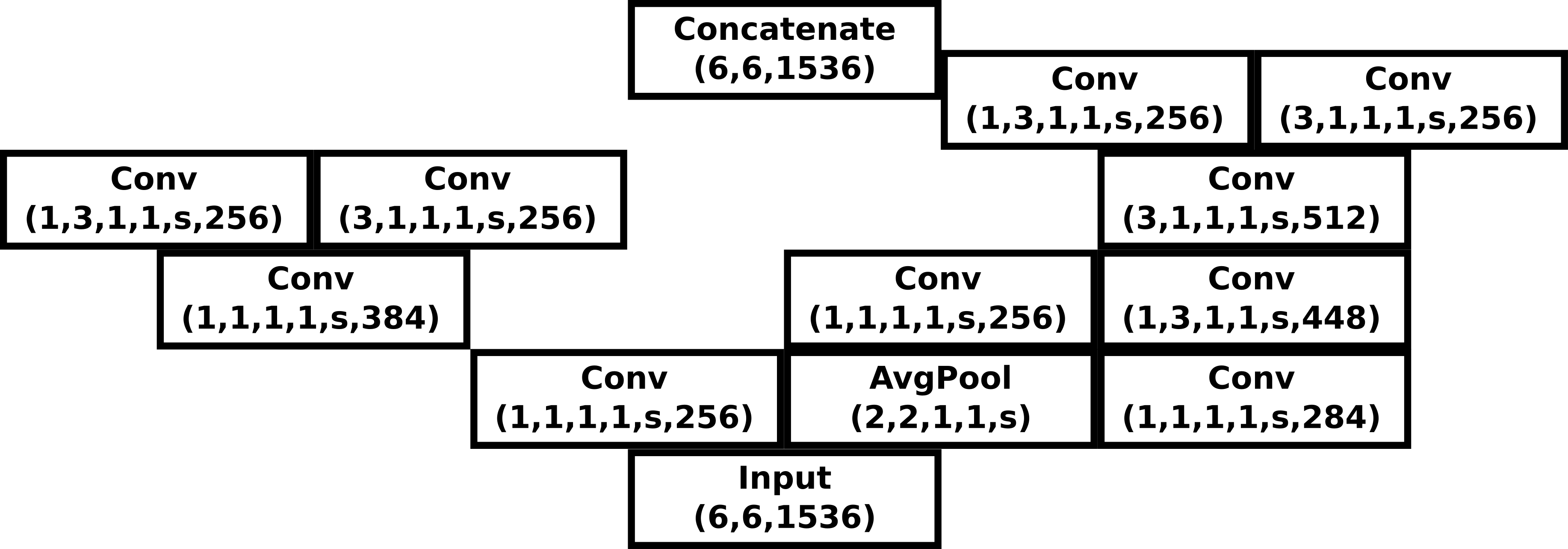}
\caption{InceptionC layer of our CNN, based on \citet{Szegedy2016}.}
\label{FIG::RESULTS::c}
\end{figure}

The ReductionA layer, shown in figure \ref{FIG::RESULTS::ra}, consists of a relatively simple combination of $3\times3$ convolutions and a MaxPooling layer. The purpose of this network module is to reduce the spatial dimension of the images from $29\times29$ pixels down to $14\times14$ in order to allow for large convolutions in the following InceptionB layer, which is shown in figure \ref{FIG::RESULTS::b}. This module consists of larger $7\times7$ convolutions, split into perpendicular stripes for runtimes reasons and hence makes an important contribution in the feature extraction process.  
Reduction layer B, shown in figure \ref{FIG::RESULTS::rb} and reduces the image dimensionality further from $14\times14$ to $6\times16$ with a rather complicated combination of convolutions. It is follow by the final InceptionC layer shown in figure \ref{FIG::RESULTS::c}, which naturally applies only small convolution but using a particularly large amount of channels.  
\begin{figure}
\includegraphics[width=.45\textwidth]{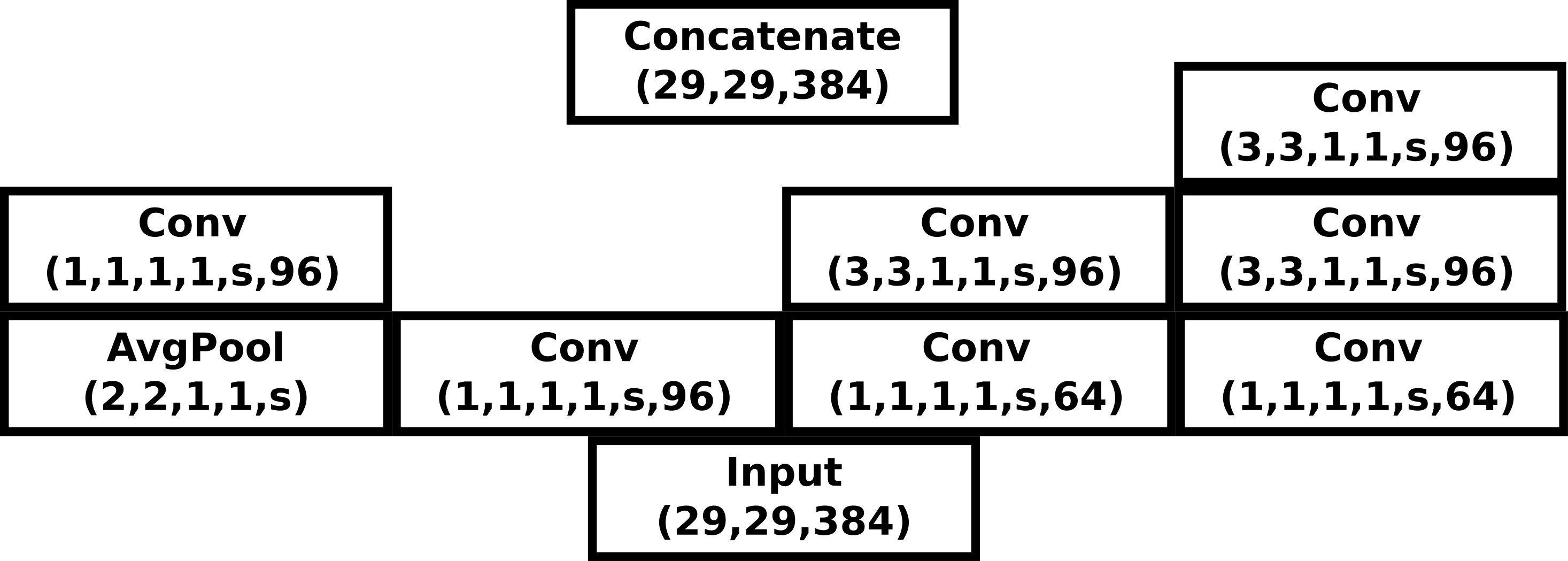}
\caption{InceptionA layer of our CNN, based on \citet{Szegedy2016}.}
\label{FIG::RESULTS::a}
\end{figure}

\bsp	
\label{lastpage}
\end{document}